\documentclass{aa}
\usepackage{graphicx}
\usepackage[varg]{txfonts}
\usepackage{natbib}
\usepackage{upgreek}
\usepackage{multirow}
\bibpunct{(}{)}{;}{a}{}{,}

\graphicspath{{images/}}

\newcommand{\M}[1]{{#1}} 

\begin{document}
   \title{Disc-binary interactions in depleted post-AGB binaries}
   \titlerunning{}
   \author{Glenn-Michael Oomen \inst{1,2}
   		   \and
           Onno Pols \inst{2} 
           \and 
           Hans Van Winckel \inst{1}
           \and
           Gijs Nelemans \inst{2,3,1}
         }
   \institute{Instituut voor Sterrenkunde (IvS), KU Leuven,
              Celestijnenlaan 200D, B-3001 Leuven, Belgium\\
              \email{glennmichael.oomen@kuleuven.be}
            \and
            Department of Astrophysics/IMAPP, Radboud University, P.O. Box 9010, 6500 GL Nijmegen, The Netherlands
            \and
            SRON, Netherlands Institute for Space Research, Sorbonnelaan 2, NL-3584 CA Utrecht, the Netherlands
   }
   \date{Received 4 May 2020 / Accepted 16 August 2020}
   \authorrunning{Oomen et al.}

\abstract{
Binary post-asymptotic giant branch (post-AGB) stars have orbital periods in the range of 100--2500 days in eccentric orbits. They are surrounded by circumbinary dusty discs. They are the immediate result of unconstrained binary interaction processes. Their observed orbital properties do not correspond to model predictions: Neither the periods nor the high eccentricities are expected. Indeed, many orbits are eccentric despite the strong tidal interaction when the primary had giant dimensions on the red giant branch (RGB) and AGB. Our goal is to investigate if interactions between a binary and its circumbinary disc during the post-AGB phase can result in their eccentric orbits, while simultaneously explaining the chemical anomaly known as depletion. For this paper, we selected three binaries (EP~Lyr, RU~Cen, HD~46703) with well-constrained orbits, luminosities, and chemical abundances. We used the \texttt{MESA} code to evolve post-AGB models, while including the accretion of metal-poor gas. This allows us to constrain the evolution of the stars and study the impact of circumbinary discs on the orbital properties of the models. We investigate the effect of torques produced by gas inside the binary cavity and the effect of Lindblad resonances on the orbit, while also including the tidal interaction following the equilibrium tide model. We find that none of our models are able to explain the high orbital eccentricities of the binaries in our sample. The accretion torque does not significantly impact the binary orbit, while Lindblad resonances can pump the eccentricity up to only $e \approx 0.2$. At higher eccentricities, the tidal interaction becomes too strong, so the high observed eccentricities cannot be reproduced. However, even if we assume tides to be ineffective, the eccentricities in our models do not exceed $\approx 0.25$. Finally, the orbit of RU~Cen is too wide to reproduce with disc-binary interactions by starting from a circular orbit. We conclude that either our knowledge of disc-binary interactions is still incomplete, or the binaries must have left their phase of strong interaction in an eccentric orbit.}

 
   \maketitle
%

\section{Introduction} \label{sect:intro}

The evolution of evolved stars in binary systems is still a poorly understood topic in astronomy \citep[for a recent review, see][]{demarco17,boffin19}. Consequently, understanding the different possible interactions between binary stars is an active field of research. In particular the interactions occurring during late stages of stellar evolution, when one of the stars has expanded to giant dimensions, still pose many uncertainties. Post-asymptotic giant branch (post-AGB) stars in binary systems are optimal candidates to study these interactions since they left their phase of strong interaction very recently. Previously, we analysed the orbital properties of a sample of 33 Galactic post-AGB binaries \citep{oomen18}. The orbital periods are in the range of 100--2500~days, with the majority having non-zero eccentricities even though tides are expected to be strong enough to circularise the orbit in the AGB phase. 

Remarkably, the orbital properties of post-AGB binaries show many similarities with other post-interaction systems, such as barium stars \citep{jorissen19,escorza19}, carbon-enhanced metal-poor stars \citep{jorissen16,hansen16}, S-type symbiotics \citep{fekel00}, blue stragglers \citep{mathieu15}, and wide subdwarf B-type (sdB) binaries \citep{vos17}. To explain the eccentric orbits in these systems, several eccentricity-pumping mechanisms have been proposed, such as Lindblad resonances in circumbinary discs \citep{dermine13,vos15,izzard18}, phase-dependent mass loss \citep{bonacicmarinovic08,kashi18}, and white-dwarf kicks \citep{izzard10}. However, binary population synthesis models are still unable to reproduce the general orbital properties of these various classes of post-interaction binaries.

The binary interaction between an AGB star and its companion leads to an epoch of strong mass loss. When most of the envelope mass of the donor is lost, the star contracts as a result of mass loss, which sparks the end of the strong interaction with its companion. At this point, the star enters the post-AGB phase, in which the star makes a transition from the AGB to a central star of a planetary nebula (CSPN) on a very short timescale \citep{vanwinckel03,millerbertolami16}. The stellar structure consists of a degenerate \element{CO} core with a mass larger than $0.51~M_\sun$, a \element{He} shell, and a low-mass (but extended) \element{H}-rich envelope of mass less than $0.02~M_\sun$. During the post-AGB phase, this remaining envelope mass is lost and the stellar radius decreases. Because this evolution happens at an almost constant luminosity, the effective temperature of the star increases from about 3000~K to 30\,000~K, at which point the central star starts to ionise its surroundings and becomes a CSPN. This evolution continues until the star becomes a hot white dwarf with $T_\mathrm{eff} \gtrsim 100\,000$~K.

A significant fraction of binaries interact on the red giant branch (RGB) instead of the AGB. The evolution of those systems can lead to the formation of the so-called post-RGB binaries \citep{kamath14,kamath15,kamath16}. The properties of those systems are very similar to those of post-AGB binaries, except for a lower luminosity ($\lesssim 2500$~$L_\sun$), and in some cases different chemical abundances due to their different nucleosynthetic history. Because post-RGB stars avoided core \element{He} burning, these stars have degenerate \element{He} cores (with $M_\mathrm{core} < 0.47$~$M_\sun$) and will evolve to eventually become \element{He} white dwarfs. Throughout this work, we will refer to the combined population of post-RGB and post-AGB stars as the post-AGB stars, unless specifically stated otherwise.

In post-AGB stars, binarity is related to the presence of a circumbinary disc, since all confirmed post-AGB binaries have a disc-type spectral energy distribution (SED) \citep{waelkens91,oomen18}. Disc-type post-AGB stars are relatively common and some 80 sources are currently known in our Galaxy \citep{deruyter06,gezer15,vanwinckel19}. Discs manifest themselves as a near- or mid-infrared excess, because hot dust is kept close to the central star. Moreover, the formation of such discs can be related to binarity, since mass loss from the AGB star will mainly be focused in the orbital plane in case of a binary. Mass flowing towards the companion leaves the system via the $L_2$ Lagrangian point, thereby creating a spiral arm \citep{frankowski07,chen17}. However, if the companion is massive enough, the spiral arm might remain bound and wrap around the binary, creating a ring as it intersects itself \citep{shu79,pejcha16,macleod18,bermudezbustamente20}. 

The presence of discs in post-AGB binaries has also been related to a chemical peculiarity called depletion \citep{vanwinckel95,vanwinckel97,maas02,kamath19}. Depletion is the systematic underabundance of refractory elements in the photospheres of post-AGB stars, in which the underabundance often scales with the condensation temperature of an element \citep{maas05,giridhar05}. The current consensus is that this depletion is caused by the accretion of metal-poor gas from the circumbinary disc \citep{waters92}. In this process of dust-gas winnowing, dust is subject to a much stronger radiation pressure than gas, hence does not get accreted. The photosphere of the post-AGB star becomes diluted with gas poor in refractory elements, which leads to the observed abundance patterns \citep{oomen19}. 

Hydrodynamical models of circumbinary discs show that accretion of gas into the binary cavity is a natural process \citep{macfadyen08, shi12, dorazio13}. The viscous evolution of the disc leads to accretion streams, taking gas from the inner rim of the circumbinary disc to the circumstellar accretion discs \citep{mosta19,munoz19a}. Most studies show that the accretion process in binaries is quite efficient, with a mass flux that is equal to or larger than in the single-star case \citep{farris14, shi15}. In a previous work \citep[][which we will refer to as paper~I]{oomen19}, we have evolved a grid of post-AGB models while accreting depleted gas at different rates. By comparing this to the observed post-AGB sample, we have shown that accretion rates of the order of $10^{-6}$~$M_\sun~\mathrm{yr}^{-1}$ are required to explain the depletion patterns of the higher-mass post-AGB population. 

In this work, we build on the results of paper~I in order to investigate the peculiar eccentric orbits of post-AGB binaries. Instead of modelling the depletion in the general population of disc-type post-AGB stars, we now focus on a more detailed approach for three specific systems. To this end, we make use of the constraints on the disc model from the depletion analysis, which provide us with important information on disc masses, accretion rates, and evolution timescales. We use this to calculate the effect of the circumbinary disc on the binary orbit. The goal is to evaluate whether current theories of disc-binary interactions are capable of reproducing both the depletion and the observed orbital eccentricities, assuming that the progenitor system leaves its strong mass-loss phase in a circular orbit at the start of the post-AGB phase. 

We discuss the different observational constraints on our stars in Sect.~\ref{sect:observationalconstraints} and we present the data on those stars in Sect.~\ref{sect:samplestars}. In Sect.~\ref{sect:methods}, we provide the methodology of our stellar evolution and binary evolution calculations, and we introduce the disc-binary interactions used in this work. We present the results in two parts. Firstly, the depletion analysis is given in Sect.~\ref{sect:depletionmodelling}. Secondly, the results of the orbital modelling are presented in Sect.~\ref{sect:orbitalmodelling}. Finally, we discuss the results in Sect.~\ref{sect:discussion} and Sect.~\ref{sect:conclusions}.

\section{Observational constraints} \label{sect:observationalconstraints}

\subsection{Luminosities} \label{sect:luminosities}
Luminosity is one of the most important parameters for post-AGB evolution. It determines the evolutionary timescale of the post-AGB phase, since it both influences the wind mass-loss rate and the nuclear burning rate, two processes that directly change the envelope mass of the star. The poorly determined luminosities of post-AGB stars are still one of the main bottlenecks in modelling their evolutionary properties. Below, we discuss two methods by which we can estimate the luminosities of the stars in our sample.

Many post-AGB stars exhibit strong radial pulsations because they are located in the Cepheid instability strip. Post-AGB stars populate the long-period end of the type~II Cepheids also known as RV~Tauri pulsators \citep{wallerstein02}. The radial modes of these stars can be directly related to their physical sizes, which together with a colour or temperature term, allows one to relate the luminosity of the star to its pulsation period. This is known as the period-luminosity-colour (PLC) relation \citep{alcock98,ripepi15}.

We use the PLC relation calibrated for RV~Tauri stars in the Large Magellanic Cloud (LMC) of \citet{manick18} to determine the luminosities of post-AGB stars in our sample. This relation uses the colour-corrected V-band magnitude known as the Wesenheit index \citep{ngeow05} and is given by
\begin{equation}
M_\mathrm{bol} = -3.75\log(P_0)+0.55+BC+2.55(V-I)_0,
\label{eq:PLC}
\end{equation}
where $P_0$ is the observed fundamental pulsation period in days, $BC$ is the bolometric correction in the $V$~band taken from \citet{flower96} based on the stellar effective temperature, and $(V-I)_0$ is the intrinsic colour of the star. The latter is determined from the (dereddened) photospheric model of the SED fit, which is constrained using spectroscopic data from the literature \citep[see][for more information]{oomen18}.

In addition to the pulsation luminosities, we can derive luminosities based on their \textit{Gaia} parallaxes \citep{gaiadr2} and the integrated photospheric fluxes of the SEDs. To do this, we use distances derived by \citet{bailerjones18} for the stars in our sample. The luminosities based on this analysis are published in paper~I for all Galactic disc-type post-AGB stars with chemical abundance data available in literature. 

However, we consider the luminosity from the pulsation analysis to be more robust, since the second data release of \textit{Gaia} does not take into account the orbital motion of the binary. This leads to systematic errors on the parallax determination, hence on the distances to our stars. Binaries with orbital periods in the range of 100--700~days are particularly prone to this effect, and often lead to an underestimation of the parallax \citep{pourbaix19}.

\subsection{Binary properties} \label{sect:binaryproperties}
Using the luminosity of the post-AGB star, it is possible to estimate its mass from the core mass-luminosity relation \citep{vassiliadis94}. For post-RGB stars, this relation is quite strict since the core mass directly determines the pressure gradient in the \element{H}-burning shell. For post-AGB stars, there is more scatter in this relation because the mass of the \element{He} shell influences the luminosity. However, it is still possible to get a good estimate for the core mass\footnote{For post-AGB stars, we will refer to the core mass as the \element{CO} core + \element{He} shell mass. For post-RGB stars, this is only the \element{He} core mass.}. Once the core mass is known, the total mass is known as well since post-AGB stars have very little mass left in their envelopes ($<0.01~M_\sun$). For the stars in our sample, we use a grid of \texttt{MESA} models (see Sect.~\ref{sect:mesamodelling}) to estimate the core masses based on the observed luminosities.

We have orbital elements available for many post-AGB stars based on a long-term, systematic, radial-velocity monitoring programme with the HERMES spectrograph \citep{raskin11} on the 1.2~m Mercator telescope. This provides us with orbital periods, eccentricities, projected semi-major axes, and mass functions \citep{oomen18,manick19}. The inclination angles are not known in general. However, based on the observed properties of the systems, it is possible to define a range of possible inclination angles in order to derive statistical distributions of the orbital properties of the binaries.

We derive several useful system properties, such as the mass of the companion ($M_2$) and the semi-major axis ($a_\mathrm{b}$). We use a statistical distribution for the inclination angle that represents random orientations in three-dimensional space given by $i = \arccos(z)$, where $z$ is the projection of the unit normal vector of the orbital plane towards the observer, which is uniformly distributed in the range of $[\cos(i_\mathrm{min}), \cos(i_\mathrm{max})]$. We derive the distribution of $M_2$ by taking 100\,000 random values from the distributions of $\sin i$, $M_1$, and $f(m)$. In a similar way, we used the resulting distribution of $M_2$ along with the distributions of $a_1 \sin i$, $\sin i$, and $M_1$ to find the distribution of $a_\mathrm{b}$. 

Finally, the Roche-lobe radius is derived using the semi-major axis and mass ratio of the system. The Roche lobe is strictly defined only for circular orbits and synchronised rotation rates. Because we are dealing with eccentric orbits, we evaluate the Roche-lobe radius at periastron using the assumption of circularity and synchronicity. Consequently, we write
\begin{equation}
R_\mathrm{L,peri} = \frac{0.49q^{2/3}}{0.6q^{2/3}+\ln(1+q^{1/3})}a_\mathrm{b}(1-e),
\label{eq:RLformula}
\end{equation}
where $R_\mathrm{L,peri}$ is the radius of a sphere with the same volume as the Roche lobe of a binary with separation equal to the periastron distance and $q$ is the mass ratio defined as $M_1/M_2$ \citep[][modified]{eggleton83}. This is a reasonable approximation that overestimates the Roche-lobe radius by up to $5\%$ \citep{sepinsky07}. If the stellar radius remains smaller than the Roche-lobe radius at periastron, the star will not fill its Roche lobe at any orbital phase, even if it is eccentric. We use the median values and 1$\sigma$ confidence intervals of the distributions of these parameters in our analysis.

\subsection{Depletion} \label{sect:depletion}

\M{In this work, we want to quantify the level of chemical depletion in post-AGB binaries by estimating the dilution of the photosphere by accreted gas.} In general, the dilution factor in the photosphere of the post-AGB star can be derived by the ratio of a volatile element, such as \element{Zn} or \element{S}, to a refractory element, such as \element{Ti} or \element{Sc}. In plateau depletion patterns, the abundances of moderate to highly refractory elements have similar abundances, such that other elements like \element{Fe} can be used together with \element{Ti}. In some cases, however, the abundances of volatile elements like \element{Zn} or \element{S} are significantly lower than those of \element{C}, \element{N}, and \element{O}. This could imply that even \element{S} and \element{Zn} are attenuated by the depletion process, hence are not suitable to measure the magnitude of depletion \citep{rao12,gezer19}. Therefore, we will use the \element{O} abundance in those cases to constrain the initial metallicity of the star, because oxygen will not be affected by the depletion process since most of the atoms are locked in \element{CO} molecules in gas phase and get accreted by the star. 

\M{Furthermore, \element{O} will not be affected much by nuclear burning processes and subsequent dredge-ups, as opposed to the \element{N} and \element{C} abundances. Indeed, it is expected that the envelope composition has been modified by the nuclear evolution of these evolved stars \citep[][and references therein]{karakas14}. However, none of the confirmed post-AGB binaries show enrichment of s-process elements \citep[with the exception of HD~158616,][]{desmedt16}. Additionally, the dust around post-AGB binaries consists of silicates \citep{gielen08, gielen11b}, although some sources exist with PAH emission and hence a carbon-rich component \citep{gielen11c}. The lack of objects with third dredge-up characteristics is possibly due to the large fraction of post-RGB stars in the sample.}

To compute the total dilution of the post-AGB star by accreted gas, we will use the average difference between the oxygen abundance on the one hand, and the iron and titanium\footnote{In case there is no \element{Ti} abundance available, \element{Sc} is used instead.} abundances on the other hand. The dilution factor of the post-AGB star then becomes
\begin{equation}
\log(\element{X_0}/\element{X}) = [\element{O}/\element{H}] - ([\element{Fe}/\element{H}]+[\element{Ti}/\element{H}])/2,
\label{eq:dilutionfactor}
\end{equation}
where $[\element{X}/\element{H}]$ is the abundance of element \element{X} with respect to the solar value in units of dex, and $\element{X_0}$ is its initial photospheric value.

\section{Sample stars} \label{sect:samplestars}

We have selected a sample of depleted post-AGB binaries based on several criteria. First, we required that the orbit of the candidate binary has been determined. The second criterion is that the chemical abundances must follow a plateau-type depletion pattern (see paper~I). \M{This allows us to better constrain the dilution factor of the photospheres for our targets. Indeed, for saturated depletion patterns we can only derive a lower limit on the dilution factor, while for plateau-type objects we can restrict successful accretion models to those that match the location of the star in the $\log(\element{X_0}/\element{X})-T_\mathrm{eff}$ plane, as discussed in Sect.~\ref{sect:mesamodelling}. The aforementioned criteria resulted in a selection of three stars}: EP~Lyr, RU~Cen, and HD~46703. In what follows, we will discuss these individually. The observational properties are given in Table~\ref{table:obscons}.

\begin{table*}
\caption{Observational constraints.}
\label{table:obscons}
\centering
\begin{tabular}{l c c c}
\hline\hline
Parameter & EP~Lyr & RU~Cen & HD~46703 \\
\hline
$P_\mathrm{orb}$ (d) & $1151\pm14$  & $1489\pm10$ & $597.4\pm0.2$ \\
Eccentricity & $0.39\pm0.09$ & $0.62\pm0.07$ & $0.30\pm0.02$ \\
$a_1 \sin i$ (AU) & $1.30\pm0.12$ & $2.38\pm0.15$ & $0.839\pm0.015$ \\
$f(m)$ ($M_\sun$) & $0.22\pm0.06$ & $0.81\pm0.17$ & $0.220\pm0.012$ \\
$L_*$ ($L_\sun$) & $5150$ & $1900$ & $2450$ \\
$T_\mathrm{eff}$ (K) & $6200\pm250$ & $6000\pm250$ & $6250\pm250$ \\
$\log(\element{X_0}/\element{X})$ (dex) & $1.87\pm0.30$ & $1.58\pm0.30$ & $1.27\pm0.30$ \\
Inclination & $[30^\circ, 60^\circ]$ & $[30^\circ, 60^\circ]$ & $[30^\circ, 60^\circ]$ \\
\hline
$M_1$ ($M_\sun$) & $0.56$ & $0.44$ & $0.46$ \\
$M_2$ ($M_\sun$) & $1.22^{+0.63}_{-0.30}$ & $2.79^{+1.94}_{-0.77}$ & $1.13^{+0.63}_{-0.24}$ \\
$a_\mathrm{b}$ (AU) & $2.68^{+0.38}_{-0.32}$ & $3.83^{+0.64}_{-0.40}$ & $1.63^{+0.19}_{-0.09}$ \\
$R_\mathrm{L,peri}$ (AU) & $0.51^{+0.11}_{-0.09}$ & $0.33\pm0.07$ & $0.35\pm0.014$ \\
\hline
\end{tabular}
\end{table*}

\subsection{EP Lyr} \label{sect:eplyr}

EP~Lyr was first classified as an RV~Tauri star by \citet{rosino51}. \citet{manick17} analysed the variability of EP~Lyr and found a fundamental pulsation period of 41.6~d, and a first overtone at 83.6~d. Based on this pulsation period, we find a luminosity of $L_\mathrm{puls} = 5150$~$L_\sun$ using Eq.~\ref{eq:PLC}. This is in agreement with the luminosity determined from the \textit{Gaia} parallax and SED fit given by $L_\mathrm{SED} = 5500^{+2300}_{-1400}$~$L_\sun$ \citep{oomen19}. This luminosity corresponds to a core mass of 0.56~$M_\sun$.

\citet{gonzalez97a} performed a spectroscopic analysis for EP~Lyr and found an (average) effective temperature of 6200~K\footnote{The effective temperature ranges from $5500$ to $7000$~K throughout the pulsation cycle, hence the formal uncertainty presented in Table~\ref{table:obscons} could be an underestimation.}. Furthermore, the chemical abundances for this star show that it is strongly depleted, and in paper~I we classified this object as having a plateau-type depletion pattern. Remarkably, the abundances of volatile elements such as \element{S} ($-0.61$~dex) and \element{Zn} ($-0.70$~dex) are significantly lower than the \element{C}, \element{N}, and \element{O} abundances ($\sim$ solar). Consequently, we derive the dilution factor from Eq.~\ref{eq:dilutionfactor}, with $[\element{Fe}/\element{H}]=-1.8$~dex and $[\element{Ti}/\element{H}]=-2.0$~dex.

The SED shows a rather weak infrared excess compared to other post-AGB stars \citep[$L_\mathrm{IR}/L_* \leq 3\%$,][]{deruyter06,gielen11b}, suggesting a low dust mass in the disc. The excess peaks at mid-infrared wavelengths, implying that the inner rim of the (dusty) disc is located far from the central binary. The cold, low-mass disc could imply that the disc is highly evolved. Furthermore, despite the oxygen-rich photosphere of the star, the disc shows a mixed chemistry with clear signs of polycyclic aromatic hydrocarbon (PAH) emission \citep{gielen11b}.

The binary nature of EP~Lyr was first discovered by \citet{gonzalez97a}. The system has a relatively wide orbit of $P_\mathrm{orb} = 1151$~days and is also significantly eccentric \citep{manick17, oomen18}. The light curve of EP~Lyr does not show any sign of periodic obscuration by a circumbinary disc \citep[known as the RVb phenomenon,][]{percy93,waelkens93,pollard96}. This implies that the inclination angle of the system is less than about 60~degrees. Furthermore, the \element{H}$\alpha$ line shows a double-peaked emission profile, which we interpret as a signature of a circumstellar accretion disc. The \element{H}$\alpha$ line has a strong blue-shifted absorption component at the orbital phase where the companion moves in front of the post-AGB star, at superior conjunction. This is indicative of a jet launched by the companion \citep{gorlova12,gorlova15,bollen17,bollen19}. Because the typical opening angle of jets in post-AGB stars is $\sim$30~degrees and systems with low inclination angles show a permanent absorption by the jet, we assume that the inclination angle of EP~Lyr is larger than 30~degrees. The orbital properties of the binary are derived based on the range of possible inclination angles of the system. The median values and 1$\sigma$ confidence intervals of the resulting distributions are presented in the lower part of Table~\ref{table:obscons}.

\subsection{RU~Cen} \label{sect:rucen}

The binary orbit of RU~Cen was first determined by \citet{gielen07} and was later updated by \citet{oomen18}. It is a wide binary with an orbital period of 1489~days and is one of the most eccentric post-AGB binaries known with $e = 0.62$. Its SED shows only a modest IR excess with $L_\mathrm{IR}/L_* = 39\%$. Furthermore, the infrared excess peaks at relatively long wavelengths ($\sim$10~$\mu$m), which implies relatively cold dust temperatures at the inner rim \citep{gielen07}, similar to EP~Lyr. Moreover, a mixed disc chemistry is observed with both oxygen- and carbon-rich dust species \citep{arneson17}.

RU~Cen has been classified as an RV~Tauri star by \citet{oconnell60}. It shows no periodic obscuration by the circumbinary disc as a result of the binary motion \citep{pollard96}. There are two dominant pulsation periods: one at 32.30~days and a first overtone at 64.64~days \citep{maas02}. This relatively short pulsation period implies a low luminosity from the PLC relation equal to $1900$~$L_\sun$. The luminosity determined from the SED fitting is lower than this value ($L_\mathrm{SED} = 1100\pm200$~$L_\sun$). Since the \textit{Gaia} parallax is unreliable for binaries, we adopt the pulsation luminosity for this object. This corresponds to a post-RGB star with a mass of $M_1 \approx 0.44$~$M_\sun$. 

A detailed chemical analysis for RU~Cen has been performed by \citet{maas02}. The effective temperature is given as $6000\pm250$~K, although the formal error of $250$~K might be underestimated given the large pulsations RU~Cen undergoes. The photosphere shows signs of strong depletion, with $[\element{Fe}/\element{H}] = -1.9$~dex and $[\element{Ti}/\element{H}] = -1.96$~dex. The \element{Zn} abundance is also relatively low at $-1$~dex, and is significantly lower that the \element{C}, \element{N}, and \element{O} abundances at around $-0.4$~dex. For that reason, we derive the dilution factor as in Eq.~\ref{eq:dilutionfactor}, which is presented in Table~\ref{table:obscons}. 

Similar to EP~Lyr, the light curve of RU~Cen does not show the RVb phenomenon, suggesting an inclination angle of less than 60~degrees. Because this is a southern target, we do not have a timeseries of optical spectra to investigate the \element{H}$\alpha$ variability. However, the mass function in this system is relatively large. If the inclination angle of RU~Cen would be less than 30~degrees, the companion mass would be larger than 7~$M_\sun$ which is very hard to explain from an evolutionary point of view. Consequently, we apply the same range in inclination angles as EP~Lyr. The resulting orbital properties are presented in Table~\ref{table:obscons}.

\subsection{HD~46703} \label{sect:hd46703}

HD~46703 was classified as a post-AGB star by \citet{luck84} based on its high luminosity and low surface gravity, and a decade later its binary nature was discovered \citep{waelkens93}. Its orbital period is about 600~days \citep{oomen18}, which is significantly smaller than that of EP~Lyr and RU~Cen. Furthermore, the orbit is significantly eccentric, with $e=0.30$. 

\citet{hrivnak08} performed a detailed photometric study of HD~46703. They classified it as a semi-regular variable, showing (relatively) small-amplitude variability ($\Delta V\approx 0.4$~mag) with a dominant pulsation period of about 29~days. Assuming that this oscillation is driven by the fundamental radial mode of the star, we derive a luminosity using Eq.~\ref{eq:PLC} equal to $2450$~$L_\sun$. The luminosity based on the \textit{Gaia} distance is much higher, namely $7500^{+3500}_{-2100}$~$L_\sun$. However, for this object, the \textit{Gaia} parallax is very likely affected by the orbital motion.

With $L_\mathrm{puls} = 2450$~$L_\sun$, HD~46703 is on the boundary of being a post-RGB or post-AGB object. The luminosity of this star is quite uncertain given the semi-regular pulsation behaviour and the discrepancy with the value based on \textit{Gaia}. Therefore, we cannot differentiate between the post-RGB or post-AGB nature of the object. Using solar-metallicity \texttt{MESA} models, we find that the luminosity of $2450~L_\sun$ is well-reproduced by a 0.46~$M_\sun$ core-mass RGB star close to igniting \element{He} in its core. Consequently, we take $M_1 = 0.46$~$M_\sun$ as the mass of the primary. 

The depletion pattern of HD~46703 shows a relatively smooth plateau profile, with similar abundances of \element{Fe} and \element{Ti} \citep{hrivnak08}. The abundances of volatile elements such as \element{S} and \element{Zn} are about 1~dex higher than those of the refractory elements, but still show an underabundance with respect to the \element{C}\element{N}\element{O} elements. Here, the same argument applies in which the \element{S} and \element{Zn} abundances could be attenuated by the depletion process.

The SED shows that there is almost no IR excess present for HD~46703, similar to EP~Lyr \citep[$L_\mathrm{IR}/L_* \leq 3\%$,][]{deruyter06,gielen11b}. The cold dust temperature implies that there is little to no dust in the disc close to the star. This could mean that the disc is either old such that the inner rim has moved away from the binary, or the dust-to-gas ratio is very small \citep{gielen11c}. The former possibility is corroborated by the low activity of the $\element{H}\alpha$ line, which implies very weak ongoing accretion. At superior conjunction, there is an (also weak) increased blue shifted absorption component. Moreover, no RVb signature is observed in the light curve. Consequently, we assume the same range of inclination angles of 30--60~degrees.

\section{Methods} \label{sect:methods}

\subsection{Overview} \label{sect:overview}
The methodology in this paper can be divided into two main parts: the depletion analysis and the binary evolution calculations. The depletion analysis is similar to that of paper~I and consists of running a grid of \texttt{MESA} models \citep{MESApaper1, MESApaper2, MESApaper3, MESApaper4} along the post-AGB phase for each of the three stars in our sample. We include accretion of depleted gas onto the post-AGB star and follow the abundance changes at the stellar surface. In doing so, we assess what accretion rates and disc masses are required to reproduce the observed depletion pattern of the star. 

\M{We simulate the evolution of the binary independently from the single-star \texttt{MESA} models in which we reproduce the depletion pattern. For the \texttt{MESA} simulations that successfully describe the depletion pattern}, we use information on the evolution of both star and disc as input for the binary evolution calculations. The disc evolution allows us to compute the effect of disc-binary interactions on the binary semi-major axis and the binary eccentricity. Finally, we assess whether at the end of the evolution, the current orbit of each post-AGB star can be adequately reproduced.

\subsection{MESA modelling} \label{sect:mesamodelling}
We used version~10398 of the one-dimensional stellar evolution code \texttt{MESA} to model the onset of chemical depletion in post-AGB stars. We evolved the post-AGB stars as single stars, but included the effect of accretion of metal-poor gas from a circumbinary disc. This method is similar to that previously used in paper~I, hence we refer the reader to Sect.~4.2 of that paper for more information on the physics used in the models.

We prepared the \texttt{MESA} models by first evolving \M{a 1.5-$M_\odot$} star along the RGB or AGB phase until the star reached the desired He core mass and by subsequently removing the envelope. For the post-RGB star RU~Cen, the luminosity suggests a core mass of 0.44~$M_\sun$ (Sect.~\ref{sect:rucen}). In this case, we evolved the star along the RGB until the helium core reached 0.44~$M_\sun$, after which we increased the mass loss rate to 10$^{-4}$~$M_\sun$~yr$^{-1}$ to remove the envelope. In order to allow for different starting points for post-RGB evolution, we remove the envelope until different amounts of envelope mass are left, or equivalently, until the star reaches different temperatures during the post-RGB phase. This method is repeated for EP~Lyr and HD~46703, but with core masses of 0.56~$M_\sun$ and 0.46~$M_\sun$, respectively.

To investigate the depletion process, we evolved a set of models with a range of values for the accretion parameters $\dot{M}_0$ and $M_\mathrm{d,0}$. These provide the time evolution of the accretion rate according to
\begin{equation}
\dot{M}(t) = \dot{M}_0\left(1+\frac{2\dot{M}_0t}{M_\mathrm{d,0}}\right)^{-3/2},
\label{eq:accroomen}
\end{equation}
which is based on a viscous disc evolution model described in detail in Appendix~\ref{sect:discmodel}. For the initial accretion rate $\dot{M}_\mathrm{0}$, we used four different values, namely $10^{-7}$, $10^{-6.6}$, $10^{-6.3}$, and $10^{-6}$ in units of $M_\sun~\mathrm{yr}^{-1}$. This range of accretion rates reproduces the global population of depleted post-AGB stars well \citep{oomen19}. For the initial disc mass $M_\mathrm{d,0}$, we used values of $10^{-3}$, $10^{-2.5}$, $10^{-2}$, $10^{-1.5}$ in units of $M_\sun$, which covers the range of observed disc masses \citep{bujarrabal13}.

The parameters $\dot{M}_\mathrm{0}$ and $M_\mathrm{d,0}$ are directly related to the viscous evolution of the disc. The disc evolves faster for higher accretion rates paired with lower disc masses, and vice versa. Therefore, the combination of $\dot{M}_\mathrm{0}$ and $M_\mathrm{d,0}$ provide constraints on the viscosity parameter $\alpha$ in the disc evolution model (see Appendix~\ref{sect:discmodel}). This information is used in the evaluation of the disc-binary interactions, discussed below. 

As was done in our previous work, we allow the post-AGB star to start its evolution at different temperatures ($T_0$). This is the temperature of the central star when the accretion from the circumbinary disc starts. This reflects the possibility that the star leaves its binary interaction with different remaining envelope masses, or equivalently, different radii. We expect the stellar radius at the start of the post-AGB phase to be determined by the Roche-lobe size. Consequently, by modelling a range of starting temperatures, we probe a range of initial Roche-lobe sizes and a range of initial semi-major axes. Possible values of $T_0$ range from the minimum temperature the star can reach on the RGB or AGB ($\sim 3000~K$) to its current effective temperature ($\sim 6000~K$). We tested different values for $T_0$ within this range and selected a region of interest for each star in the sample in order to fully probe the parameter space of the accretion model (see Sect.~\ref{sect:depletionmodelling}).

Next, we select those models that can reproduce the observed depletion pattern of each post-AGB star in the $\log(\element{X_0}/\element{X})-T_\mathrm{eff}$ plane. This in turn allows us to determine the evolutionary timescale of the post-AGB star from the start of accretion until the current point in evolution. Furthermore, the time evolution of the accretion rate (Eq.~\ref{eq:accroomen}), and hence of the disc mass, are also known. This information is used to evaluate the interactions between the disc and the binary which we describe in the next subsection.

\subsection{Disc-binary interactions} \label{sect:disc-binaryinteractions}
There are several ways in which the disc can interact with the binary. Firstly, the binary can induce density waves at resonant locations in the disc, which carry energy and angular momentum from the binary to the disc \citep{lubow96, dermine13, vos15}. This mechanism decreases the orbital angular momentum of the binary. Secondly, accretion of high-angular momentum gas onto the binary acts to increase the total angular momentum of the binary \citep{munoz19a, moody19, munoz19b, duffell19}. In this case, gas inside the binary cavity exerts a positive gravitational torque on both stars.

\subsubsection{Resonance torque} \label{sect:resonancetorque}
Smoothed-particle hydrodynamics (SPH) simulations by \citet{artymowicz91} show that the rotating gravitational potential of a binary can exert a significant torque on the disc by producing spiral density waves at resonant locations in the disc \citep{goldreich79}. Its effect on eccentricity in binary evolution models has been investigated in detail by \citet{dermine13} and \citet{vos15}, while \citet{rafikov16b} includes the effects of disc evolution in his models. An important characteristic of the resonance torque is that it is independent of the strength and the width of the resonance area, but is completely determined by the global evolution of the disc \citep{lubow96}. In other words, one can write
\begin{equation}
\dot{J}_\mathrm{res} = -J_\mathrm{d}/\tau_{\nu} = -J_\mathrm{d} \alpha \left(\frac{H}{R}\right)^2 \Omega(R),
\label{eq:jdotres}
\end{equation}
where $J_\mathrm{d}$ is the total angular momentum of the circumbinary disc, \M{$\tau_{\nu} = (\alpha \left(\frac{H}{R}\right)^2 \Omega(R))^{-1}$} is the (global) viscous evolution timescale of the disc, $(H/R)$ is the aspect ratio of the disc, and $\Omega(R)$ is the Keplerian rotation rate at the half-angular momentum radius $R$. The viscosity parameter $\alpha$ is constrained by the accretion parameters and the binary properties, as is described in Appendix~\ref{sect:discmodel}. Typical values of $\alpha$ found in literature for protoplanetary discs range from $10^{-3}$--$10^{-1}$ \citep[e.g.][]{andrews09,rafikov17,ansdell18}, which largely agrees with values we find from our depletion analysis (see final column in Tables~\ref{table:resultshd46703}--\ref{table:resultseplyr}). In Eq.~\ref{eq:jdotres}, we use the viscous timescale as defined by \citet{lubow96}, since the resonance torque in their SPH simulations agrees well with this definition. \M{We assume a thick disc with aspect ratio $(H/R) = 0.2$ \citep{kluska18}, which develops the high disc luminosities generally observed in post-AGB binaries.}  

On the basis of SPH simulations, it is shown that at small and moderate eccentricities the outer Lindblad resonance at $l=1$ and $m=2$ is most important \citep{artymowicz91}. This resonance is located at $\Omega = \Omega_\mathrm{b}/3$ and is responsible for truncating the inner rim of the disc \citep{lubow96}. At small or moderate eccentricities ($e \leq 0.2$), we describe the change in eccentricity from the $(l,m)=(1,2)$ resonance as \citep{dermine13}
\begin{equation}
\dot{e}_\mathrm{res} = \frac{l}{m}\frac{2}{\tau_{\nu}}\frac{J_\mathrm{d}}{J_\mathrm{orb}}\frac{1-e^2}{e+\frac{0.01\alpha}{e}}\left(\frac{1}{\sqrt{1-e^2}}-\frac{l}{m}\right).
\label{eq:edotlowe}
\end{equation}
Here, $J_\mathrm{orb}$ is the \M{orbital angular momentum of the binary and $\tau_{\nu}$ is as defined in Eq.~\ref{eq:jdotres}}. The change in eccentricity is strongest at low eccentricities and reaches its maximum when $e \simeq 0.1\alpha^{1/2}$. However, below this critical point, the $(l,m)=(1,2)$ resonance can no longer support the inner rim of the disc, and $\dot{e}$ rapidly vanishes to zero \citep{lubow96}. At eccentricities larger than this critical point, $\dot{e}$ decreases as $1/e$. This trend continues as the eccentricity increases and more resonances become excited. When the eccentricity reaches $0.5-0.7$, these resonances start to cancel, effectively reducing $\dot{e}$ to zero. In order to model the change in eccentricity in the regime $0.2 < e \leq 0.7$, we follow \citet{vos15} by using
\begin{equation}
\dot{e}_\mathrm{res} = \frac{l}{m}\frac{2}{\tau_{\nu}}\frac{J_\mathrm{d}}{J_\mathrm{orb}}\left(\frac{c_1}{e} + c_2\right)
\label{eq:edothighe}
\end{equation}
where $c_1$ and $c_2$ are constants that are computed such that Eq.~\ref{eq:edothighe} smoothly connects to Eq.~\ref{eq:edotlowe} at $e=0.2$, while going to zero at $e=0.7$. When the eccentricity increases beyond 0.7, we set $\dot{e}_\mathrm{res}$ equal to zero.

\subsubsection{Accretion torque} \label{sect:accretiontorque}
Recent hydrodynamical simulations have shown that gas inside the circumbinary disc cavity exerts a torque on the binary \citep{miranda17}. \citet{munoz19a} show that the bulk of the torque arises from the extended circumstellar discs around both components of the binary. The overall torque that impacts the binary scales linearly with the mass accretion rate and the specific orbital angular momentum of the binary. Other recent studies performing 2D and 3D simulations of circumbinary discs also show that the binary gains angular momentum from disc interactions \citep{munoz19a, moody19, munoz19b, duffell19}. There is a remarkable agreement between these different works and different system parameters on the proportion of this relation, which is well-described by
\begin{equation}
\dot{J}_\mathrm{accr} = 0.7\dot{M}a_\mathrm{b}^2\Omega_\mathrm{b},
\label{eq:jdotaccr}
\end{equation}
where $\dot{J}_\mathrm{accr}$ is the torque due to accretion, $\dot{M}$ is the accretion rate onto the binary, \M{$a_\mathrm{b}$ is the binary semi-major axis,} and $\Omega_\mathrm{b}$ is the orbital angular velocity.

\citet{munoz19a} also compute the change in eccentricity in their models by evaluating the forces exerted by each cell in the simulation on the components of the binary for four different eccentricities, namely $0.0$, $0.1$, $0.5$, $0.6$. Even though the torques are similar in each of these models, the effect on the eccentricity is different. For the circular case, the magnitude of $\dot{e}$ is close to zero, while for $e = 0.1$ the eccentricity increases, and for the large-eccentricity models the eccentricity decreases. This suggests that the sign of $\dot{e}$ flips at some point (near $e \approx 0.5$). However, since there are only three relevant data points available, we cannot derive a reliable analytic formulation of $\dot{e}$ as a result of accretion. In order to see whether this mechanism can produce the eccentric orbits observed in binary post-AGB stars, we will assume that the effect of accretion is independent of eccentricity and has the largest value found in the simulations of \citet{munoz19a} (at $e=0.1$) which is 
\begin{equation}
\dot{e}_\mathrm{accr} = 2.42 \dot{M} M_\mathrm{b}^{-1}. 
\label{eq:eaccr}
\end{equation}

In the models by \citet{munoz19a, moody19, munoz19b, duffell19}, the Lindblad resonances as described by \citet{lubow96} do not seem to be excited, or produce a much smaller torque than predicted by the model. \citet{munoz19a} show that the torque produced by the circumbinary disc is about a factor of 20 smaller than the torque produced by the gas inside the cavity. However, from an order of magnitude estimation (taking $\tau_\nu \approx 0.5M_\mathrm{d}/\dot{M}$, $J_\mathrm{orb} \approx M_\mathrm{b}a_\mathrm{b}^2 \Omega_\mathrm{b}$, and assuming the specific angular momentum of the disc is about twice that of the binary orbit), one finds that the torques produced by resonances in the disc should be of the same order or even larger than those from gas accretion. We interpret this as implying either that the resonances are much weaker in reality than predicted by the analytical model of \citet{lubow96}, or that grid-based hydrodynamical codes cannot properly model the resonance torques, as opposed to the more ballistic approach of the SPH code of \citet{artymowicz91}.

\subsection{Binary evolution modelling} \label{sect:binaryevolutionmodelling}
The depletion analysis using \texttt{MESA} modelling gives information on the combinations of starting temperature ($T_0$), initial disc mass ($M_\mathrm{d,0}$), and initial accretion rate ($\dot{M}_0$) that can explain the depletion pattern. For each of the successful accretion models, we use this information in the binary evolution calculations. \M{These binary evolution calculations are completely decoupled from the single-star \texttt{MESA} models discussed in Sect.~\ref{sect:mesamodelling} to maximise computational efficiency.} We present the methodology of the binary evolution modelling below.

The first step is selecting an initial orbit for the binary. We always start from an almost circular orbit ($e = 0.01$). To select the initial semi-major axis $a_\mathrm{b,0}$ of the binary, we impose two boundary conditions. \M{The first condition is that we require the star to remain inside its Roche lobe at all times in the evolution. Since we treat the depletion analysis in \texttt{MESA} and the binary evolution separately, we cannot include the effect of Roche-lobe overflow in our simulations. This condition is verified by comparing the Roche-lobe radius (Eq.~\ref{eq:RLformula}) to the stellar radius of the \texttt{MESA} model at each time step. If a model for a given $a_\mathrm{b,0}$ experiences Roche-lobe overflow at any point in its evolution, the initial orbit is likely too compact and the model is rejected.}

\M{The second condition we impose is that the initial semi-major axis must be small enough such that Roche-lobe overflow (RLOF) could have occurred at some point in its previous evolution. Post-RGB systems can only form if an external agent (i.e. the companion) removes its envelope on the RGB, since stellar winds are not strong enough at this point. We impose this condition by computing the maximum initial semi-major axis based on the maximum possible radius a star can reach for its given core mass while varying the envelope mass. HD~46703 has a core mass of about 0.46~$M_\sun$. The largest size a star with this core mass can have on the RGB is approximately 0.8~AU. From Eq.~\ref{eq:RLformula}, the star would fill its Roche lobe in a circular orbit for $a_\mathrm{b,0} \approx 2.7$~AU (using the current masses of both components). Hence we impose a maximum $a_\mathrm{b,0}$ of 2.7~AU, such that the progenitor of HD~46703 can still have filled its Roche lobe in the past. Applying a similar approach to RU~Cen, we find $R_\mathrm{max} \approx 0.7$~AU and $a_\mathrm{max} \approx 3.0$~AU, which is larger than for HD~46703 because its companion mass is higher. For EP~Lyr, we have have a maximum radius on the AGB of $R_\mathrm{max} \approx 1.3$~AU and we find $a_\mathrm{max} \approx 4.5$~AU. Though this restriction is not strictly relevant for the progeny of AGB stars, we keep this assumption for the post-AGB models of EP~Lyr since disc formation is often associated with the substantial overfilling of the donor star's Roche lobe. Regardless, we find that the condition of $a_\mathrm{b,0} \leq 4.5$~AU does not restrict the parameter space of EP~Lyr in our results below.}

At the start of the binary evolution calculations, we initialise the initial disc parameters discussed in Appendix~\ref{sect:discmodel}. The initial inner and outer disc radii are determined based on the initial semi-major axis of the system. Next, the disc angular momentum is computed by integrating from the inner to the outer radius. Once the initial conditions for both disc and binary have been set, we evolve the system up to the current point in evolution, which is determined by the depletion analysis. 

In each time step, we compute the relevant torques acting on the disc and binary and evaluate the changes in angular momentum, semi-major axis, and eccentricity. We use a simple forward Euler method to integrate the parameters ($a_\mathrm{b}$, $e$, and $J_\mathrm{d}$) over the evolution timescale defined by the depletion modelling. \M{We divide the evolution timescale into equal time steps for the integration of the orbit. Our tests show that 10\,000~time steps are necessary and sufficient to ensure a smooth and accurate evolution, while keeping the computational cost relatively low. Using only $10^3$~time steps gives irregularities for the slowly evolving post-RGB models due to the non-linear behaviour of the resonance and tidal torques, while increasing the number of time steps to $10^5$ produces almost identical results as $10^4$ time steps.}

The initial semi-major axis is then optimised using a least-squares minimalisation to find the model for which the final orbit is as close as possible to the observed orbit (in $e$ vs $a_\mathrm{b}$ space). Even though the uncertainties in the observations in Table~\ref{table:obscons} do not have a Gaussian distribution, the least-squares method will provide a good indication of whether the disc-binary interactions can adequately reproduce the orbit.

Several mechanisms act to change the orbital parameters of the binary. The disc-binary interactions discussed in Sect.~\ref{sect:disc-binaryinteractions} produce torques which result in changes in eccentricity and semi-major axis. Additionally, an important binary process we consider is spin-orbit coupling due to tides \citep{zahn77}. The differential gravitational force of the companion causes distortions in the equilibrium shape of the star. The companion exerts a torque on the post-AGB star due to the dissipation of energy in the tidal bulges, driving the rotational spin to synchronicity with the orbital frequency. Furthermore, the loss of mechanical energy due to tidal friction acts to reduce the eccentricity in most circumstances \citep{hut81}. The details on how we handle tides in our models are described in Appendix~\ref{sect:equilibriumtide}. 

To evaluate the effect of the disc-binary interactions on the orbit, we investigate several scenarios, presented in Table~\ref{table:cases}, in which we start the evolution from a circular orbit ($e=0.01$) and allow disc-binary interactions to pump the eccentricity. The first case is our standard scenario, in which we include the effects of resonances, accretion, and tides on the binary orbit. This will be discussed intensively in Sect.~\ref{sect:case1results}. In the next two cases, we simulate the evolution while ignoring the torque from accretion (case~2) and resonances (case~3). This allows us to better assess the effects of these mechanisms on the orbit. Finally, we investigate the possibility that the equilibrium tide model is not adequate for post-AGB stars and that tides are not effective. In this case, we remove the effect of tides on the binary.

Based on these different cases, we compute the total torque acting on the binary at each time step as
\begin{equation}
\dot{J}_\mathrm{orb} = \dot{J}_\mathrm{accr} + \dot{J}_\mathrm{res} + \dot{J}_\mathrm{tides},
\label{eq:jdottot}
\end{equation}
where $\dot{J}_\mathrm{accr}$ is given by Eq.~\ref{eq:jdotaccr}, and $\dot{J}_\mathrm{res}$ is given by Eq.~\ref{eq:jdotres}. The torque originating from tides follows from spin-orbit coupling, hence $\dot{J}_\mathrm{tides} = -\dot{J}_\mathrm{tides,s}$ with $\dot{J}_\mathrm{tides,s}$ given by Eq.~\ref{eq:tidesspin}\footnote{Angular momentum terms related to spin are indicated by an extra subscript $s$.}. For the disc, we can write $\dot{J}_\mathrm{d} = - \dot{J}_\mathrm{res} - \dot{J}_\mathrm{accr}$, which are the negative contributions of the disc-binary terms in Eq.~\ref{eq:jdottot}. For each case, we only take into account the relevant terms as described in Table~\ref{table:cases}. \M{We note that the torques contributing to Eq.~\ref{eq:jdottot} are averaged over the orbit, hence it is unnecessary to treat these equations as phase dependent.} 

Similarly, we can write the total effect on the eccentricity as
\begin{equation}
\dot{e} = \dot{e}_\mathrm{accr} + \dot{e}_\mathrm{res} + \dot{e}_\mathrm{tides},
\label{eq:edottot}
\end{equation}
where $\dot{e}_\mathrm{accr}$ is as computed from Eq.~\ref{eq:eaccr} and $\dot{e}_\mathrm{res}$ is as given in Sect.~\ref{sect:resonancetorque}. The effect of tides is given in Appendix~\ref{sect:equilibriumtide} and provides either a positive or negative contribution to $\dot{e}$, depending on the rotational velocity of the star. 

Finally, we use Eqs.~\ref{eq:jdottot} and \ref{eq:edottot} to compute the change in semi-major axis by using 
\begin{equation}
\dot{a}_\mathrm{b} = 2a_\mathrm{b}\left(\frac{\dot{J}_\mathrm{orb}}{J_\mathrm{orb}} + \frac{e\dot{e}}{(1-e^2)}\right).
\label{eq:aevol}
\end{equation}
Here we assume that the masses of both stars are conserved. This is a fair assumption, since $M_\mathrm{b} \gg M_\mathrm{d}$.

\begin{table}
\renewcommand{\arraystretch}{1.5}
\caption{Different cases investigated in this work.}
\label{table:cases}
\centering
\begin{tabular}{l c c c}
\hline\hline
Cases & Tides & Resonances & Accretion\\
\hline
Case~1 & $\times$ & $\times$ & $\times$ \\
Case~2 & $\times$ & $\times$ & \\
Case~3 & $\times$ & & $\times$ \\
Case~4 & & $\times$ & $\times$ \\
\hline
\end{tabular}
\end{table}

\section{Results: depletion modelling} \label{sect:depletionmodelling}

\subsection{HD~46703} \label{sect:hd46703results}

\begin{figure}
\resizebox{\hsize}{!}{\includegraphics{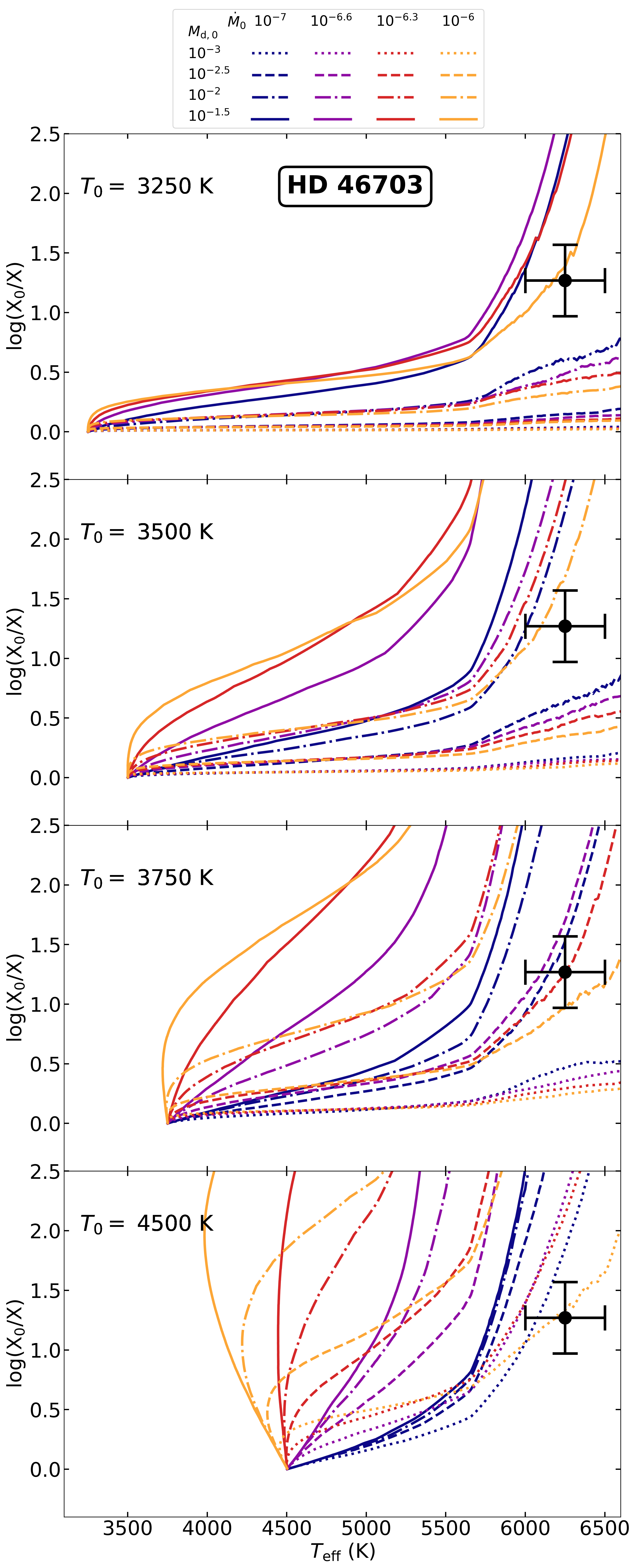}}
\caption{Depletion versus $T_\mathrm{eff}$ plot for HD~46703 with starting temperatures of 3250~K, 3500~K, 3750~K, and 4500~K from \textit{top} to \textit{bottom}, respectively. The black error bar denotes the observed value of HD~46703. The legend is given at the top of the figure, in which different colours and line styles represent different initial accretion rates and initial disc masses, respectively. $\dot{M}_0$ is expressed in $M_\sun~\mathrm{yr}^{-1}$ and $M_\mathrm{d,0}$ is given in $M_\sun$.}
\label{fig:hd46703depletion}
\end{figure}

We selected a range of starting temperatures ($T_0$) to cover the parameter space of the accretion model as well as possible. For HD~46703, this results in four different starting temperatures: 3250~K, 3500~K, 3750~K, and 4500~K. At $T_0 < 3250$~K, the models evolve too slowly and do not become depleted even for the highest disc masses and accretion rates, while at $T_0 > 4500$~K, our lowest accretion rates are still too high and the models are depleted too quickly. Models for different $T_0$ are depicted in four separate panels in Fig.~\ref{fig:hd46703depletion}. Because HD~46703 has a plateau-type depletion pattern, successful models are required to go through the observed location in the $\log(\element{X_0}/\element{X})-T_\mathrm{eff}$ plane. All of the models that do cross the 1$\sigma$ boundary covered by the observations are used in the analysis of the disc binary interactions below and are presented in Table~\ref{table:accretionparams}.  

In Fig.~\ref{fig:hd46703depletion}, there is a clear anti-correlation between $T_0$ and initial disc mass. This trend is explained by the fact that the stellar atmosphere can be diluted most rapidly at high $T_\mathrm{eff}$ values, because at that point the outer convective envelope mass becomes very low. Consequently, what matters most is the rate of accretion at those high $T_\mathrm{eff}$. In slowly evolving post-RGB stars, like HD~46703 and RU~Cen, the evolution timescale is long, such that the main parameter determining the accretion rate is the initial disc mass \citep[see Fig.~5 in][]{oomen19}. 

We see that for the lowest $T_0$ of 3250~K (upper panel of Fig.~\ref{fig:hd46703depletion}), only the highest disc-mass models are capable of reproducing the depletion pattern of HD~46703. The star still has a substantial \element{H}-rich envelope at 3250~K of $\sim0.03$~$M_\sun$, hence it takes over 100\,000~years before it reaches its current point in evolution. In order for the disc to still provide sufficient accretion at later stages in the evolution, the initial disc mass should be high.

\subsection{RU~Cen} \label{sect:rucenresults}

\begin{figure}[!]
\resizebox{\hsize}{!}{\includegraphics{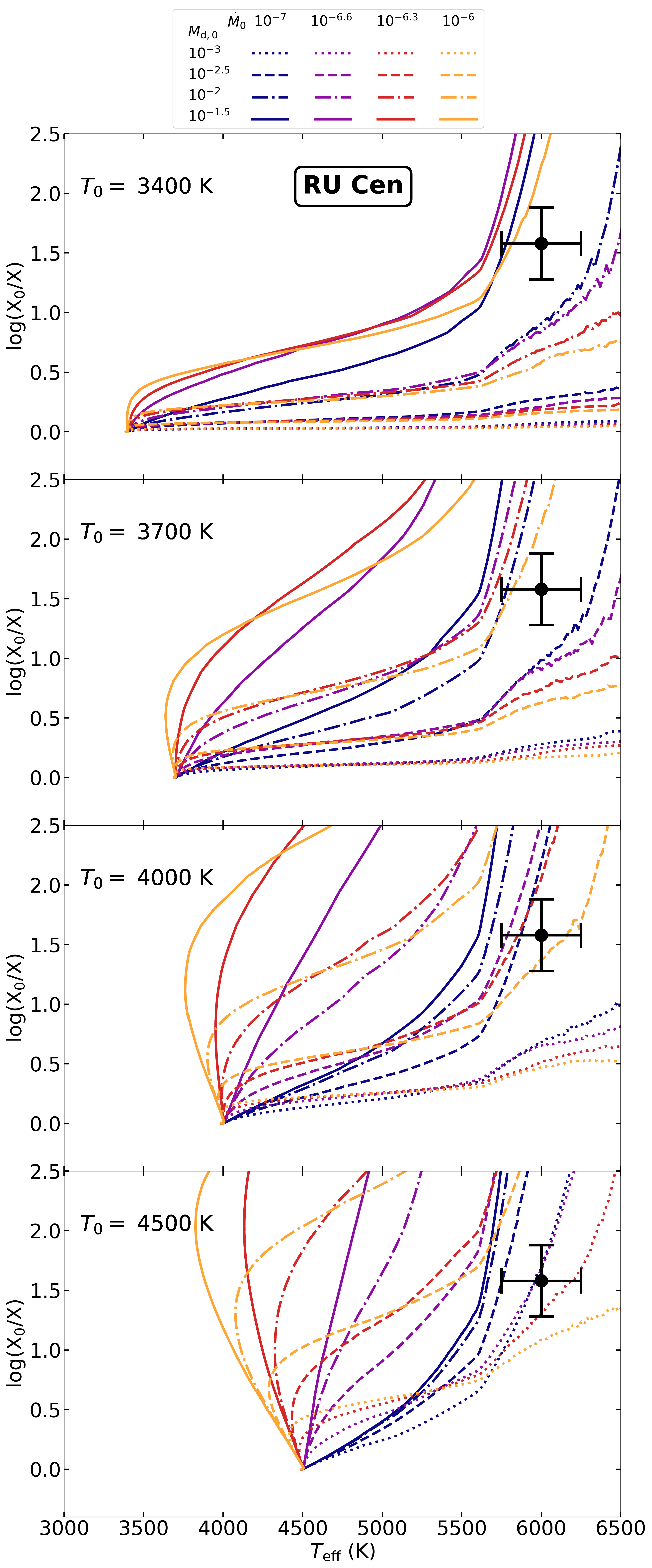}}
\caption{Depletion versus $T_\mathrm{eff}$ plot for RU~Cen with starting temperatures of 3400~K, 3700~K, 4000~K, and 4500~K from \textit{top} to \textit{bottom}, respectively. See legend at the top for definitions of colours and line styles.}
\label{fig:rucendepletion}
\end{figure}

For RU~Cen, the full parameter space of the accretion model is well represented by models with $T_0$ values of 3400~K, 3700~K, 4000~K, and 4500~K (see Fig.~\ref{fig:rucendepletion} from top to bottom, respectively). Similar to HD~46703, RU~Cen is a slowly evolving post-RGB star of low luminosity. Consequently, we observe the same trend in which lower values of $T_0$ require higher initial disc masses to become depleted. The collection of models that fit the observed location of RU~Cen are presented in Table~\ref{table:accretionparams}.

\subsection{EP~Lyr} \label{sect:eplyrresults}
\begin{figure}
\resizebox{\hsize}{!}{\includegraphics{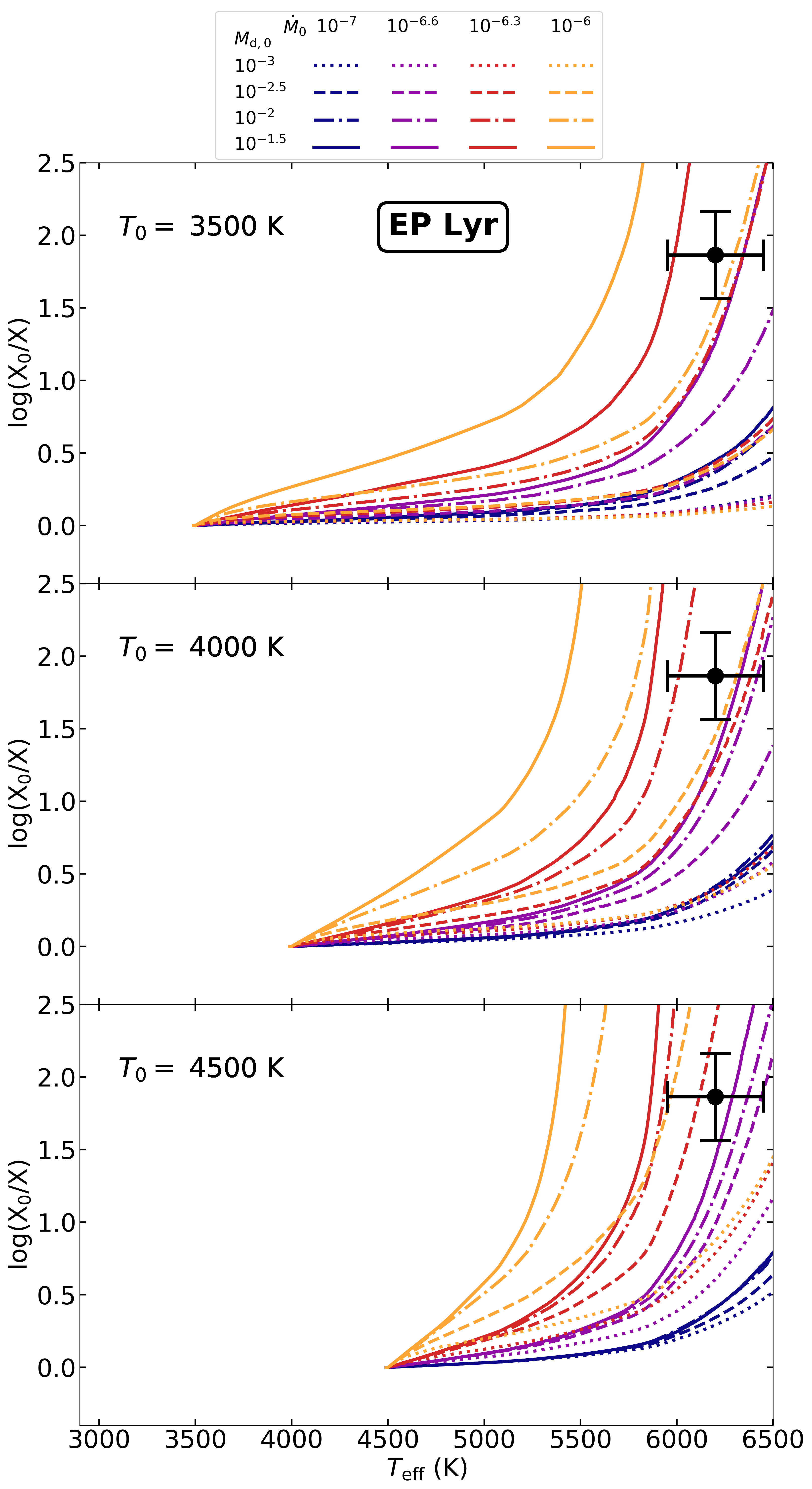}}
\caption{Depletion versus $T_\mathrm{eff}$ plots for EP~Lyr with starting temperatures of 3500~K, 4000~K, and 4500~K from \textit{top} to \textit{bottom}, respectively. See legend at the top for definitions of colours and line styles.}
\label{fig:eplyrdepletion}
\end{figure}

For EP~Lyr, we selected three values for $T_0$: 3500~K, 4000~K, and 4500~K. These models are depicted in three separate panels in Fig.~\ref{fig:eplyrdepletion}. Increasing $T_0$ beyond 4500~K produces similar results as those of $T_0 = 4500$~K. We therefore restrict our parameter space to the given range.

For EP~Lyr, the set of models that fit the observed parameters in Fig.~\ref{fig:eplyrdepletion} are more heterogeneous in terms of disc mass compared to the other two stars in our sample. This illustrates the degeneracy in the parameter space between $\dot{M}_0$ and $M_\mathrm{d,0}$. However, we find that in none of the cases, models with low accretion rates of $\dot{M}_0 = 10^{-7}$~$M_\sun~\mathrm{yr}^{-1}$ are able to reproduce the observed depletion pattern of EP~Lyr. This is caused by the fast evolution of a 0.56~$M_\sun$ post-AGB star, as was shown in paper~I.

\section{Results: orbital modelling} \label{sect:orbitalmodelling}

\subsection{Case~1} \label{sect:case1results}

\begin{figure}
\resizebox{\hsize}{!}{\includegraphics{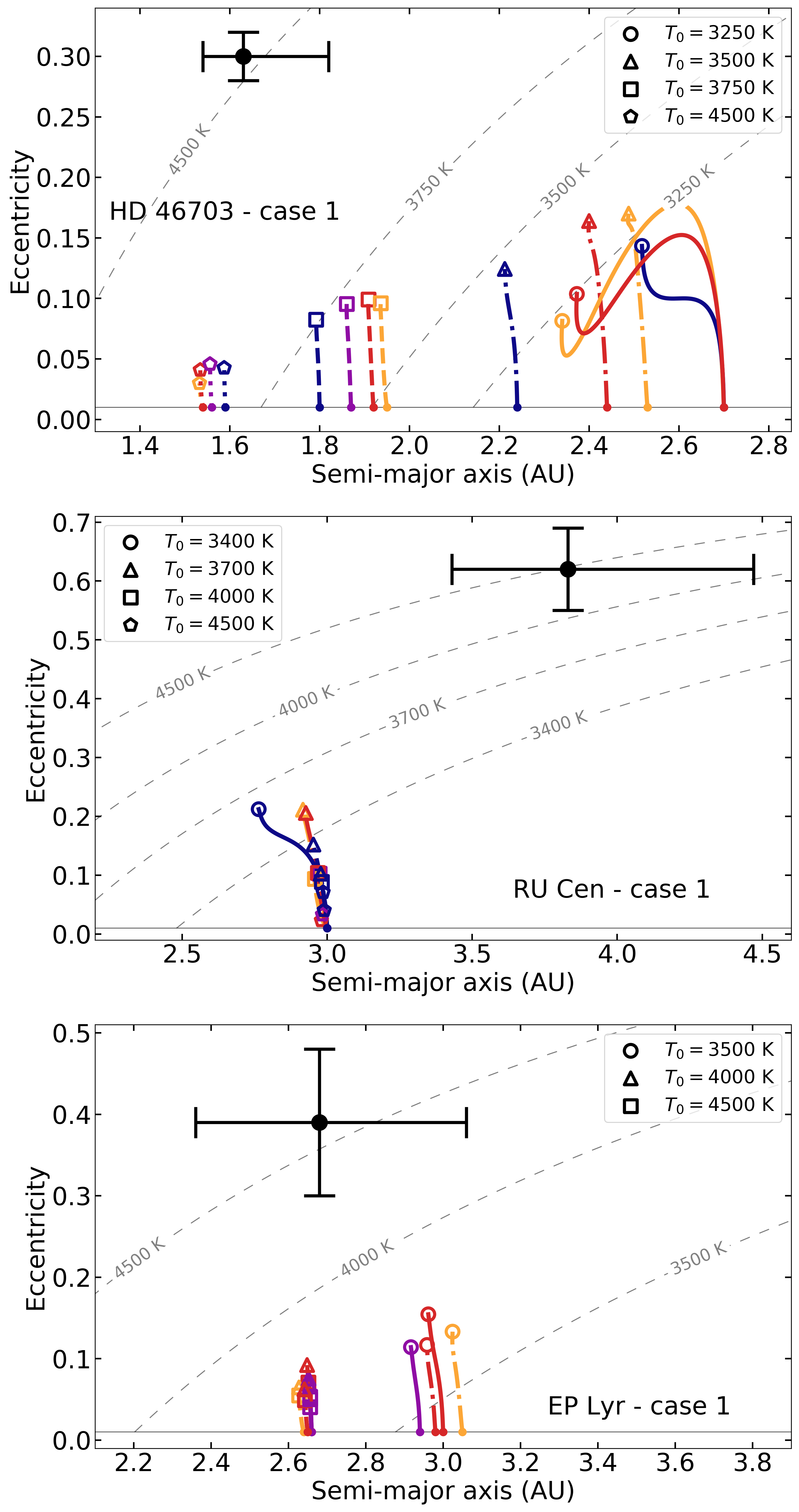}}
\caption{Best-fitting models in the $e$~vs~$a_\mathrm{b}$ plane for case~1 with both accretion and resonance torques. The black error bars show the location for each of the stars in the sample. The line styles used in this plot are the same as for the models that fit the observations in Figs.~\ref{fig:hd46703depletion}--\ref{fig:eplyrdepletion}. The symbols at the end points represent models of different $T_0$. The models start from a circular orbit with $e=0.01$, which is shown by the thin black line. The grey dashed curves are lines of constant $R_\mathrm{L}$, in which the given temperature corresponds to the radius for this star's luminosity such that $R_* = R_\mathrm{L}$. Models for HD~46703 are shown in the \textit{top panel}, for RU~Cen in the \textit{middle panel}, and for EP~Lyr in the \textit{bottom panel}.}
\label{fig:case1}
\end{figure}

In case~1, we investigate the combined effect of the accretion torque and the resonance torque on the evolution of the binary. The accretion torque is exerted by gas inside the disc cavity and leads to an increase in orbital angular momentum, while the resonance torque removes orbital angular momentum by launching spiral density waves at the inner rim of the circumbinary disc. As discussed in Sect.~\ref{sect:disc-binaryinteractions}, both mechanisms pump the binary eccentricity and change the semi-major axis. Furthermore, we also take into account the tidal interaction, which tends to circularise the binary. 

Figure~\ref{fig:case1} shows the evolution of the binary orbit in the $e$~vs~$a_\mathrm{b}$ plane for each of the stars in the sample. In each of the panels, we show the best-fitting models (by least-squares analysis) for all the accretion models that reproduce the observed depletion value in Figs.~\ref{fig:hd46703depletion}--\ref{fig:eplyrdepletion}. The models start from a circular orbit ($e=0.01$) and gradually become eccentric due to the interaction with the disc. However, none of the models can reproduce the orbital properties of the stars in our sample. The initial and final orbital parameters for all the models are shown in Tables~\ref{table:resultshd46703}--\ref{table:resultseplyr}.

The top panel of Fig.~\ref{fig:case1} shows the case-1 models for HD~46703. Several trends can be observed here. \M{The low-$T_0$ models start and end in wider orbits.} This is in the first place because the lower $T_0$ models are limited by their Roche lobes. The grey dashed curves in Fig.~\ref{fig:hd46703case2maxa} show lines of constant $R_\mathrm{L}$, such that models above those lines will fill their Roche lobes at that given temperature. As a star evolves, it becomes hotter and is able to evolve above those lines without filling its Roche lobe. Additionally, models with lower $T_\mathrm{eff}$ have a larger $R_*$ and are subject to much stronger tidal forces, since $\dot{e}_\mathrm{tides} \propto (R_*/a_\mathrm{b})^8$. Consequently, these models need to start in wider orbits so as to reduce the tidal interaction. For the models with $T_0 = 3250$~K in the top panel of Fig.~\ref{fig:case1}, the effect of tides is particularly clear, as the eccentricity decreases at later times in the evolution.

\begin{figure}
\resizebox{\hsize}{!}{\includegraphics{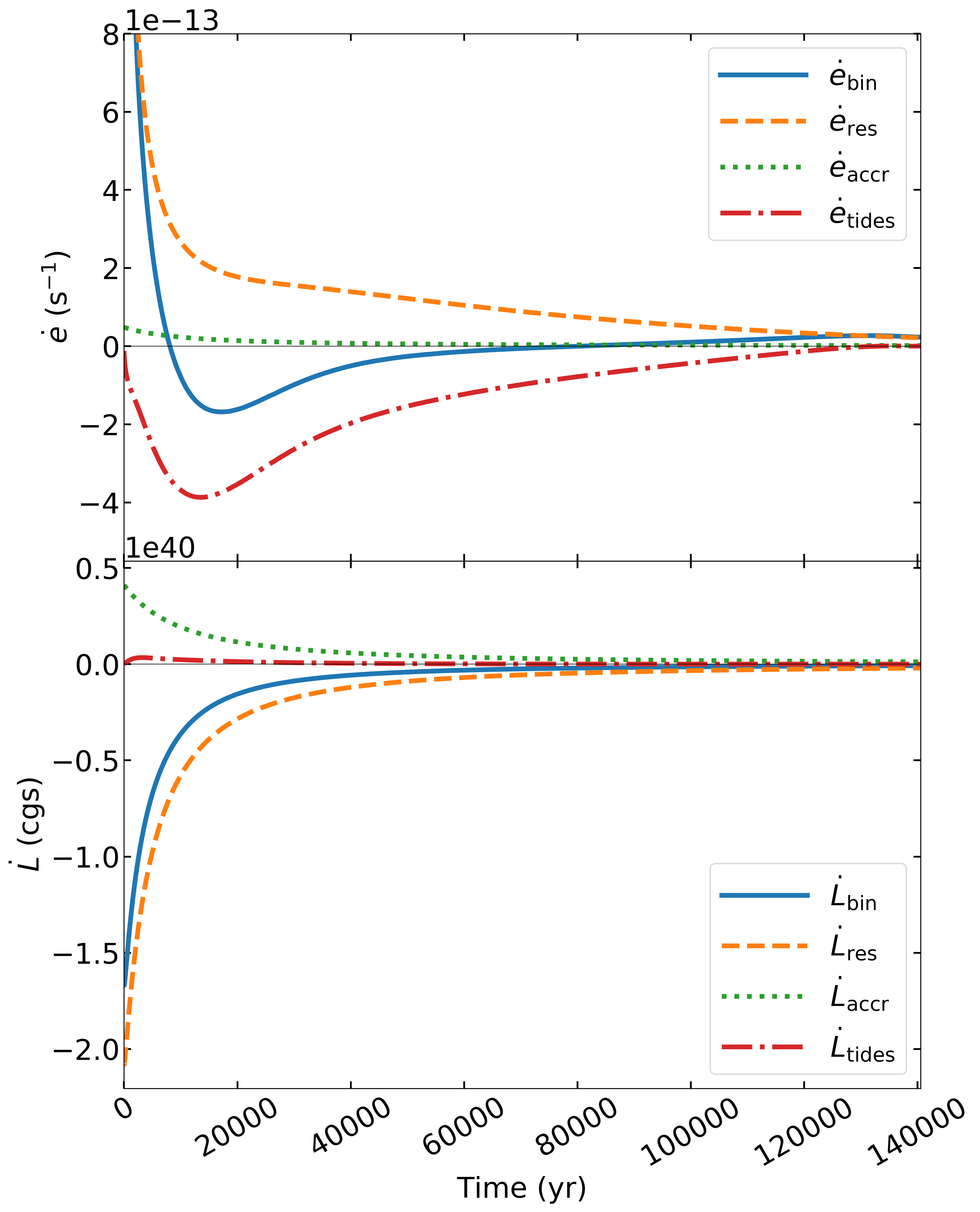}}
\caption{Time evolution of $\dot{e}$ (\textit{top panel}) and $\dot{L}$ (\textit{bottom panel}) for the different mechanisms used in case~1 for HD~46703. The parameters used in this model are $M_\mathrm{d,0} = 10^{-1.5}$~$M_\sun$ and $\dot{M}_0 = 10^{-6.0}$~$M_\sun~\mathrm{yr}^{-1}$ with $T_0 = 3250$~K, which corresponds to the solid orange line in the top panel of Fig.~\ref{fig:case1}. The solid blue line represents the sum of the three other effects, i.e. resonances, accretion, and tides. The two most important effects on the orbital evolution are resonances and tides.}
\label{fig:hd46703figs}
\end{figure}

To evaluate the effect of the different mechanisms acting on the eccentricity, we show the evolution of $\dot{e}$ for a model of HD~46703 with parameters $T_0 = 3250$~K, $\dot{M} = 10^{-6}$~$M_\sun~\mathrm{yr}^{-1}$, and $M_\mathrm{d,0} = 10^{-1.5}$~$M_\sun$ (solid orange line in top panel of Fig.~\ref{fig:case1}) in the top panel of Fig.~\ref{fig:hd46703figs}. The two most important effects on the orbital eccentricity are tides and resonances. Accretion only has a minute effect on the eccentricity. This can be understood from the small eccentricity pumping effect in Eq.~\ref{eq:eaccr}. Since the total disc mass is in general less than $1\%$ of the total binary mass, the maximum pumping of eccentricity will be a $\sim 0.02$ increase.

Resonances are most effective at low eccentricities, which results in a large $\dot{e}_\mathrm{bin}$ early in the evolution. However, as the eccentricity increases, tides become stronger since $\dot{e}_\mathrm{tides}$ scales linearly with eccentricity to first order. This effect is countered, however, by the contraction of the star. Since $\dot{e}_\mathrm{tides} \propto (R_*/a_\mathrm{b})^8$, the tides reduce in strength again as the star evolves, and almost vanish when the effective temperature of the star increases beyond 5000~K. This allows $e$ to increase again slightly.

The evolution of the torque exerted on the binary system is shown in the bottom panel of Fig.~\ref{fig:hd46703figs}. Overall, the resonance torque is about five times stronger than the accretion torque throughout the evolution of the disc. The net result is that the binary loses angular momentum, resulting in the decrease of the semi-major axis for HD~46703 in Fig.~\ref{fig:case1}. Furthermore, it is clear from Fig.~\ref{fig:hd46703figs} that the spin-orbit coupling due to tides only weakly changes $J_\mathrm{orb}$, but provides a positive contribution since the star rotates super-synchronously during its evolution (see Fig.~\ref{fig:corotation}).


We show the case-1 models for RU~Cen in the middle panel of Fig.~\ref{fig:case1}. The maximum initial semi-major axis for RU~Cen to have experienced Roche-lobe overflow is $a_\mathrm{max} = 3.0$~AU, as described in Sect.~\ref{sect:binaryevolutionmodelling}. This semi-major axis is smaller than the currently observed semi-major axis of 3.8~AU (albeit in an eccentric orbit). This causes the most optimal models for RU~Cen to start at the maximum $a_\mathrm{b,0}$. The final semi-major axes of the models are smaller than the observed value and the eccentricities reach values slightly higher than 0.2.

To reproduce the orbit of RU~Cen, the semi-major axis should increase, while the eccentricity needs to be pumped substantially to its currently observed value of $0.62\pm0.07$. While the resonance torque is the most effective interaction to pump the eccentricity, this can only lead to a decrease of the orbital size. It is therefore impossible for this mechanism to produce the current orbital properties of RU~Cen. Moreover, the large eccentricity of 0.62 is hard to explain for any orbital size using resonances, since these will start to cancel when the eccentricity reaches $\sim0.5$ and become very ineffective \citep{artymowicz91}.


The bottom panel of Fig.~\ref{fig:case1} displays the case-1 models of EP~Lyr. The high luminosity of EP~Lyr results in a faster evolution for this object, with timescales in the range of 1000--10\,000~yr (see Table~\ref{table:accretionparams}). The disc-binary interactions do not have enough time to pump the eccentricity up to values above 0.15. Models with $T_0 = 4000$~K and $T_0 = 4500$~K have a final semi-major axis close to the observed value, while models with $T_0 = 3500$~K start in wider orbits to avoid Roche-lobe overflow throughout their evolution.

To better understand the effects of the individual disc-binary interaction mechanisms, we show the evolution of the models with only resonance torque (case~2) and only accretion torque (case~3) in the next subsection. We focus on HD~46703 since this star shows the greatest diversity in possible evolution scenarios. Finally, we will discuss models in which we ignore the equilibrium tide (case~4) in Sect.~\ref{sect:case4results}.

\subsection{Case~2 and case~3} \label{sect:case2and3results}

\begin{figure}
\resizebox{\hsize}{!}{\includegraphics{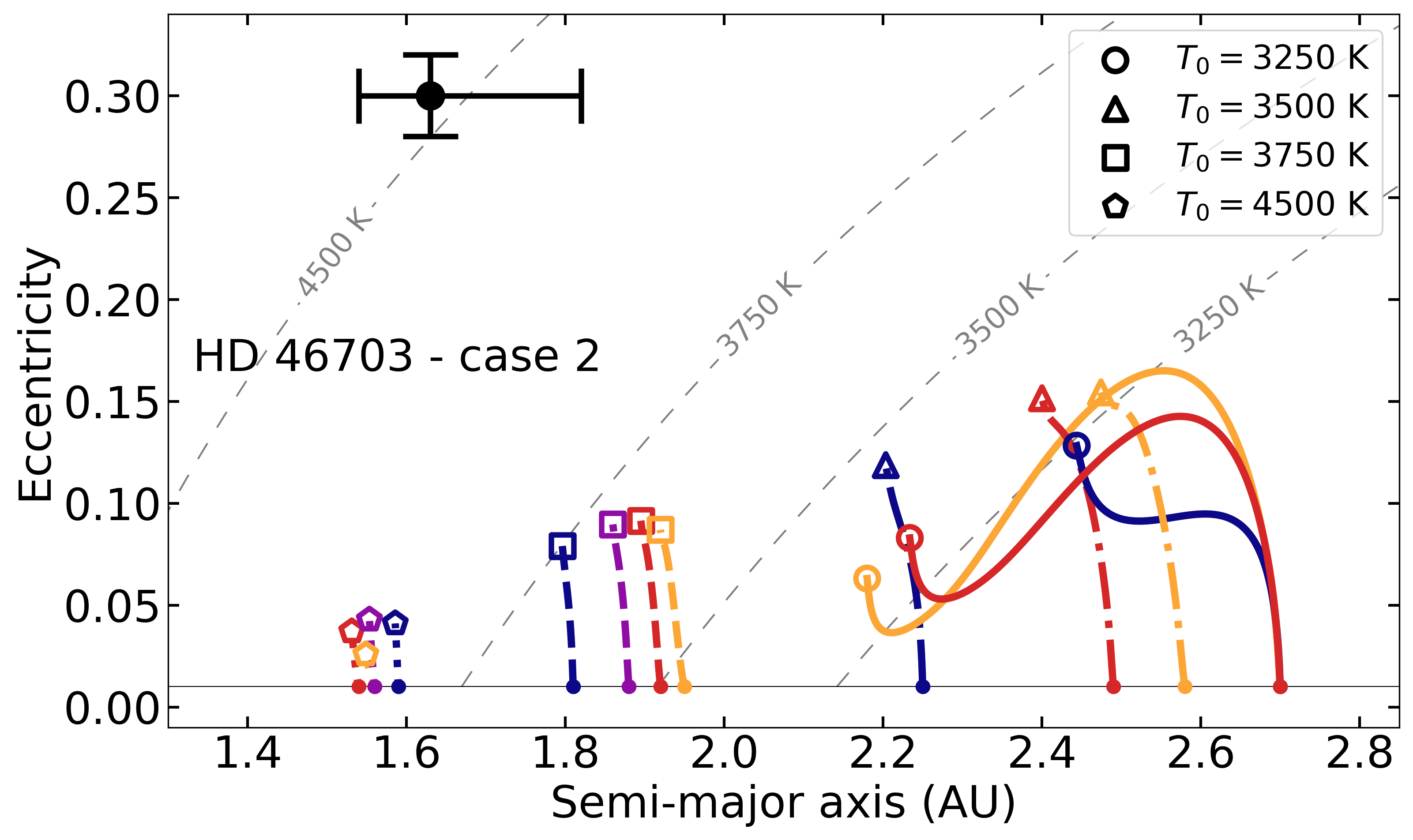}}
\caption{Best-fitting models of HD~46703 in the $e$~vs~$a_\mathrm{b}$ plane for the effect of resonances in case~2. The symbols and line styles are the same as in the upper panel of Fig.~\ref{fig:case1}.}
\label{fig:hd46703case2maxa}
\end{figure}

In case~2, we analyse the effect on the orbit when we only include the resonance torque described in Sect.~\ref{sect:resonancetorque}, along with the tidal interaction. Lindblad resonances drive spiral density waves at the inner rim of the disc, which take away orbital energy and angular momentum from the binary. Figure~\ref{fig:hd46703case2maxa} shows the evolution of the models for HD~46703 in the $e$~vs~$a_\mathrm{b}$ plane. The model results are presented in Table~\ref{table:resultshd46703}.

It is clear that the results for case~2 are very similar to those of case~1. As was shown in Fig.~\ref{fig:hd46703figs}, the resonance torque has a much larger impact on the orbit than the accretion torque. The Lindblad resonances are effective at pumping eccentricities, as was already shown by \citet{dermine13,vos15}. However, for the case of HD~46703, the eccentricities only increase to $\sim 0.15$, which is far below the observed value of $0.30 \pm 0.02$. Moreover, the final semi-major axis is too large for many of the low-$T_0$ models. Consequently, we cannot simultaneously reproduce the depletion pattern and the orbit of HD~46703.

\begin{figure}
\resizebox{\hsize}{!}{\includegraphics{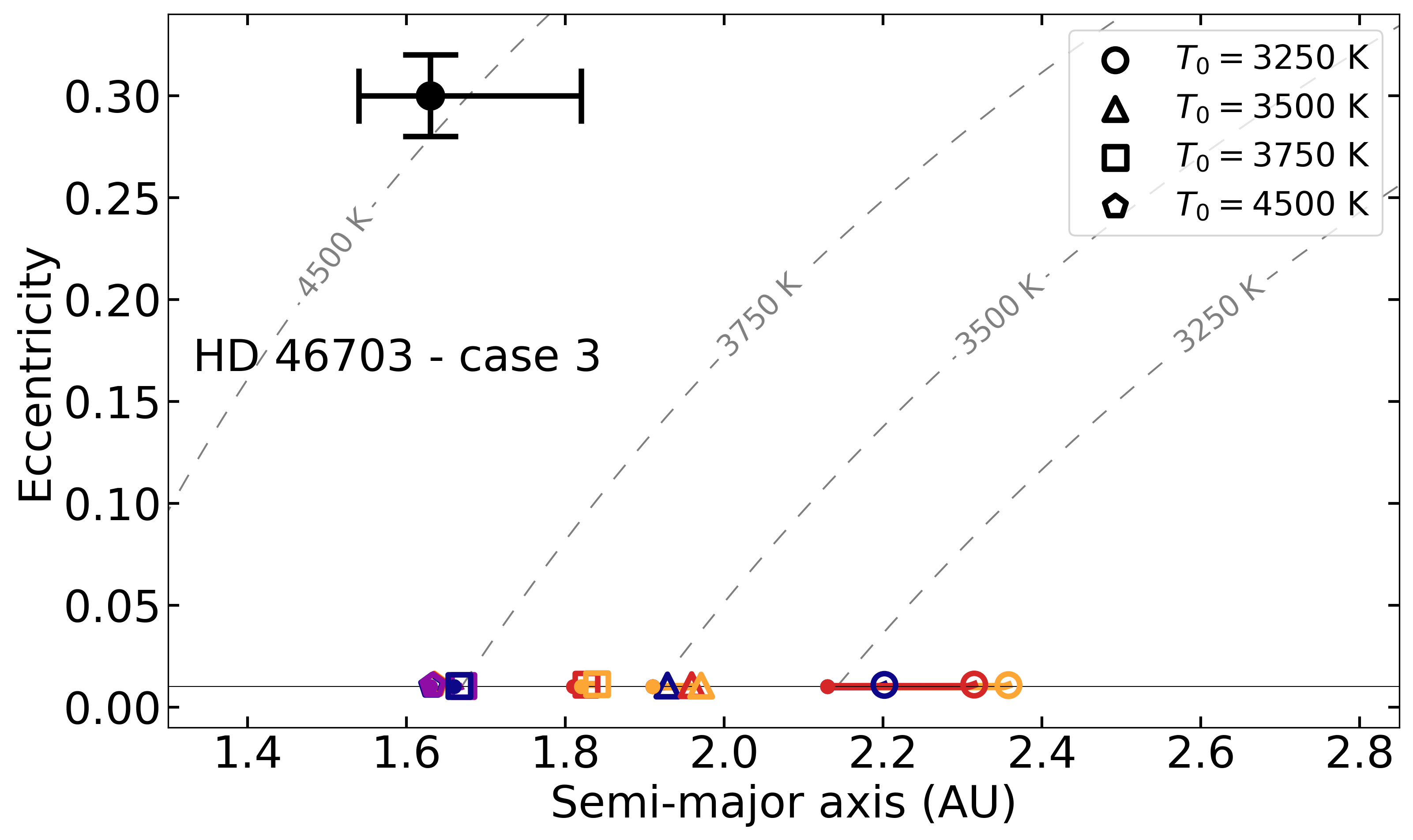}}
\caption{Same as Fig.~\ref{fig:hd46703case2maxa}, but for the accretion torque described in Sect.~\ref{sect:accretiontorque}.}
\label{fig:hd46703case3}
\end{figure}

In case~3, we investigate the effect of the accretion torque presented in Sect.~\ref{sect:accretiontorque}. This exerts a positive torque on the binary including an eccentricity pumping effect. The orbital evolution is shown for the case of HD~46703 in Fig.~\ref{fig:hd46703case3} and presented in Table~\ref{table:resultshd46703}. Accretion results in modest widening of the orbit for the low-$T_0$ models. All of the models remain circular, which confirms the small effect of accretion on the eccentricity, as shown in Fig.~\ref{fig:hd46703figs}. We note that we assumed the maximum eccentricity increase found in the models of \citet{munoz19a}; the actual effect is likely to be even smaller.

\subsection{Case~4} \label{sect:case4results}
\begin{figure}
\resizebox{\hsize}{!}{\includegraphics{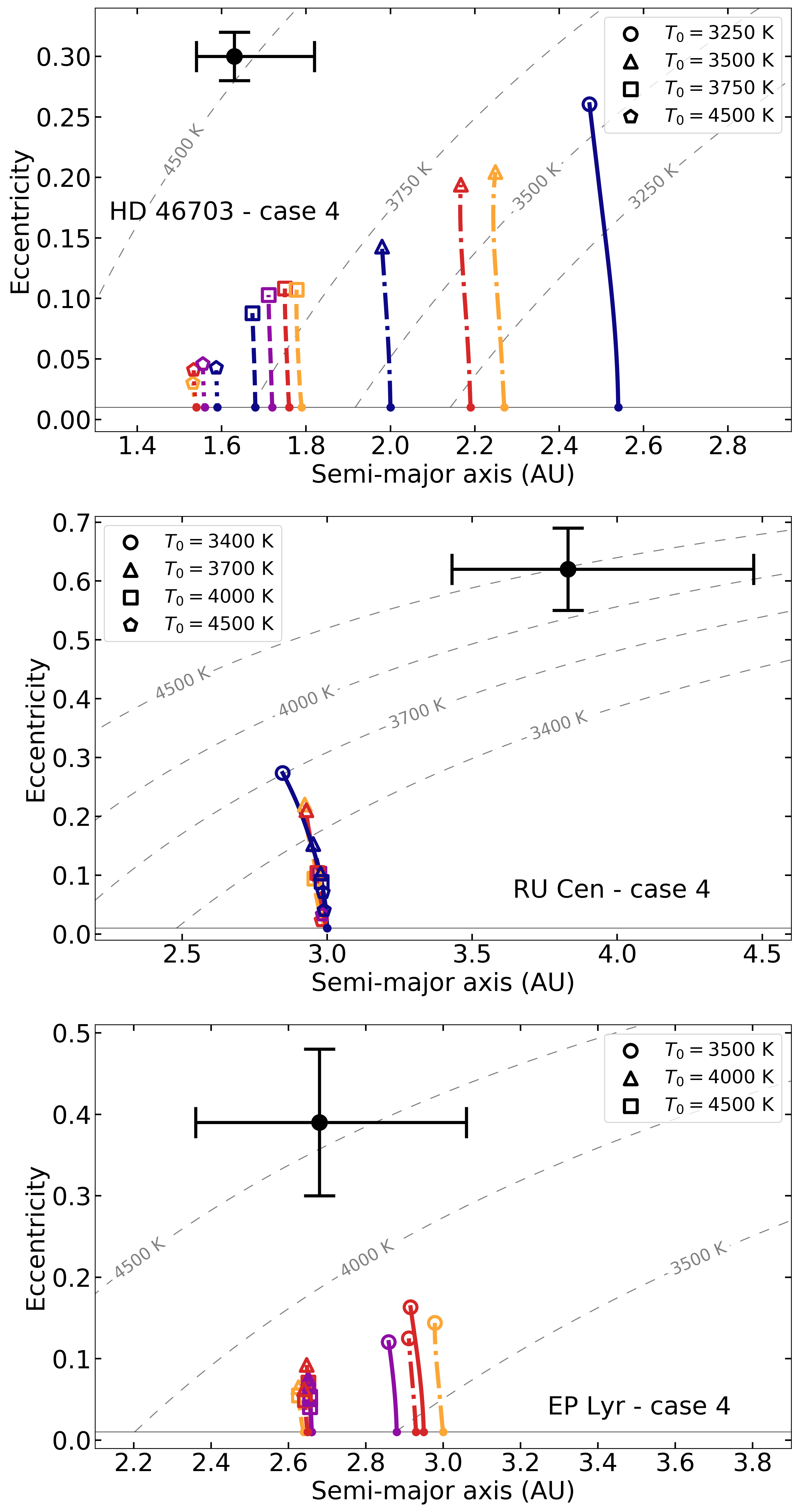}}
\caption{Same as Fig.~\ref{fig:case1}, but for models in which the effect of tides is ignored (case~4).}
\label{fig:case4}
\end{figure}

Case~4 explores the scenario in which tides are ignored, as a way of simulating the possibility that tides are not as strong as theory predicts. The equilibrium tide theory is well tested for the case of a red-giant star with a large, extended envelope. Even though post-AGB stars have an extended red-giant like structure, the envelope has a very low mass and density and convective damping may be less effective than for a normal red giant. Moreover, post-AGB envelopes are in a state of non-homologous contraction. In Fig.~\ref{fig:case4}, we show the results of models with resonance and accretion torques, but without the effect of tides. 

In the case of HD~46703, we observe two main differences in Fig.~\ref{fig:case4} with respect to the models that include tides in Fig.~\ref{fig:case1}. Firstly, the models reach higher eccentricities, with the low-$T_0$ models coming close to the observed eccentricity. Secondly, we see a general shift of the models to lower $a_\mathrm{b,0}$ compared to case~1. The latter is due to the strong \M{dependence of tides on the semi-major axis}. Models for case~1 starting at larger semi-major axes are able to become more eccentric, hence provide a better fit to the observed value. The model with $T_0 = 3250$~K in the top panel of Fig~\ref{fig:case4} reaches $e \approx 0.25$, which is close to the observed value of 0.30. However, the final semi-major axis is much larger than the observed value. The other two models of $T_0 = 3250$~K with higher accretion rate (solid orange and red lines in Fig.~\ref{fig:case1}) end up filling their Roche lobes for case~4, even when starting at $a_\mathrm{max} = 2.7$~AU. The general result of case~4 for HD~46703 is that the models cannot simultaneously explain both orbital size and eccentricity.

For RU~Cen and EP~Lyr, the results of case~4 in Fig.~\ref{fig:case4} are similar to those of case~1. This is because tides play a less important role in the orbital evolution of these systems. RU~Cen is still limited by the maximum initial semi-major axis of the system. For EP~Lyr, the evolution timescale is too short for the eccentricity to be pumped to significant values by the disc-binary interactions. The parameters for all case-4 models are presented in Tables~\ref{table:resultshd46703}--\ref{table:resultseplyr}.

\subsection{Assumptions} \label{sect:assumptions}
In this work, we have made several assumptions within our models. Here, we investigate the effect of some of these assumptions on the results derived above.

\subsubsection{Initial outer disc radius} \label{sect:initialouterdiscradius}
In our disc model (Appendix~\ref{sect:discmodel}), we assumed that the initial outer disc radius $R_\mathrm{out,0}$ is situated at $10a_\mathrm{b}$ in order to calculate the initial disc angular momentum. However, since both $R_\mathrm{out,0}$ and the initial disc angular momentum are unknown physical quantities, we investigate its effect on our results by changing its value for several models. 

We arbitrarily chose to test this assumption for models of HD~46703 with $T_0=3750$~K. We recalculate the models for case~1, but change the value of $R_\mathrm{out,0}$ to $5a_\mathrm{b}$ and $50a_\mathrm{b}$. The results are given at the bottom of Table~\ref{table:resultshd46703}. We find that the final results are similar to those of the standard model. Consequently, the value of the initial outer disc radius has little impact on the orbital evolution, and our assumed value of $10a_\mathrm{b}$ is reasonable.

The effect of this parameter can be understood from its interaction with the resonance torque, which has the largest effect on the binary properties in our models. Equation~\ref{eq:jdotres} gives the strength of the torque exerted by resonances as the ratio of $J_\mathrm{d}/\tau_\nu$. Increasing the value of $R_\mathrm{out,0}$ increases the initial disc angular momentum via Eq.~\ref{eq:angmomint}. Even though this also results in a larger viscosity parameter (Eq.~\ref{eq:viscoustimescale}), the overall viscous timescale increases since the half-angular momentum radius scales linearly with $R_\mathrm{out,0}$ which significantly reduces $\Omega(R)$. Consequently, the overall effect of increasing the outer radius is a small reduction in the resonance torque, which does not significantly impact our results. On the other hand, changing $R_\mathrm{out,0}$ does result in changes in the viscosity parameter $\alpha$. Since neither variable is directly constrained by observations for our objects, we cannot make conclusive statements on the values of these parameters based on our models.

\subsubsection{Disc scale height} \label{sect:discscaleheight}

\M{In our binary evolution models, we assumed a thick disc model with a large aspect ratio $(H/R) = 0.2$. This value is based on the large scale height inferred from post-AGB discs \citep{kluska18}. Furthermore, when comparing the viscous timescale from the resonance model of \citet{lubow96} to that of our disc model in Eq.~\ref{eq:viscoustimescale}, we find that these timescales are similar when taking $(H/R) \approx 0.2$. As already noted in the previous subsection, the choice of $R_\mathrm{out,0}$ also influences the initial viscous timescale.}
	
\M{In the resonance model of \citet{lubow96}, the global viscous timescale of the disc is related to the scale height via the unknown viscosity parameter $\alpha$. By constraining the aspect ratio in our models, we therefore parametrise the viscous timescale via the viscosity parameter. Since we fix the viscosity parameter by our depletion analysis in \texttt{MESA}, one can see that a larger value for the aspect ratio represents a shorter viscous timescale of the disc in the resonance model, hence results in a stronger torque and more eccentricity pumping.}

\M{In an attempt to quantify the influence of the disc aspect ratio on our results, we evaluated the effect of a smaller value of this ratio, namely $(H/R) = 0.1$, which might be more representative for objects with a smaller infrared excess, such as EP~Lyr and HD~46703. We present the results for models of HD~46703 with $T_0=3750$~K at the bottom of Table~\ref{table:resultshd46703}. The effect is quite clear: a lower aspect ratio results in less eccentricity pumping. The final eccentricity of these models is about half of that of our standard models.}

\subsubsection{Stellar wind strength} \label{sect:stellarwindstrength}

\begin{figure}
\resizebox{\hsize}{!}{\includegraphics{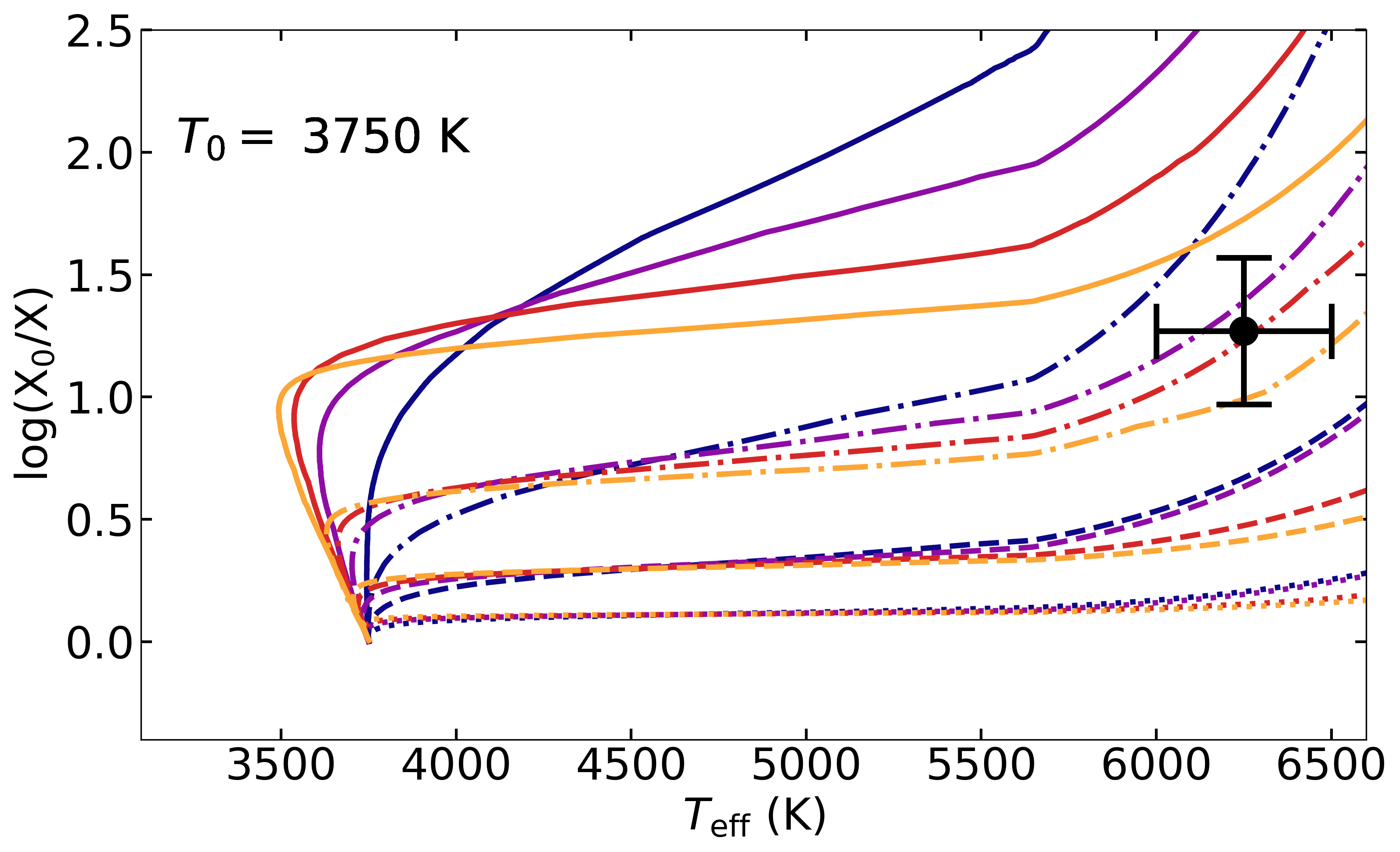}}
\caption{Same as the bottom panel of Fig.~\ref{fig:hd46703depletion} (with $T_0 = 3750$~K), but for models with a stellar wind which is a factor of 100 weaker. In this case, more massive discs are required to explain the observed depletion pattern due to the longer evolution timescale.}
\label{fig:hd46703_weakwind}
\end{figure}

\begin{figure}
\resizebox{\hsize}{!}{\includegraphics{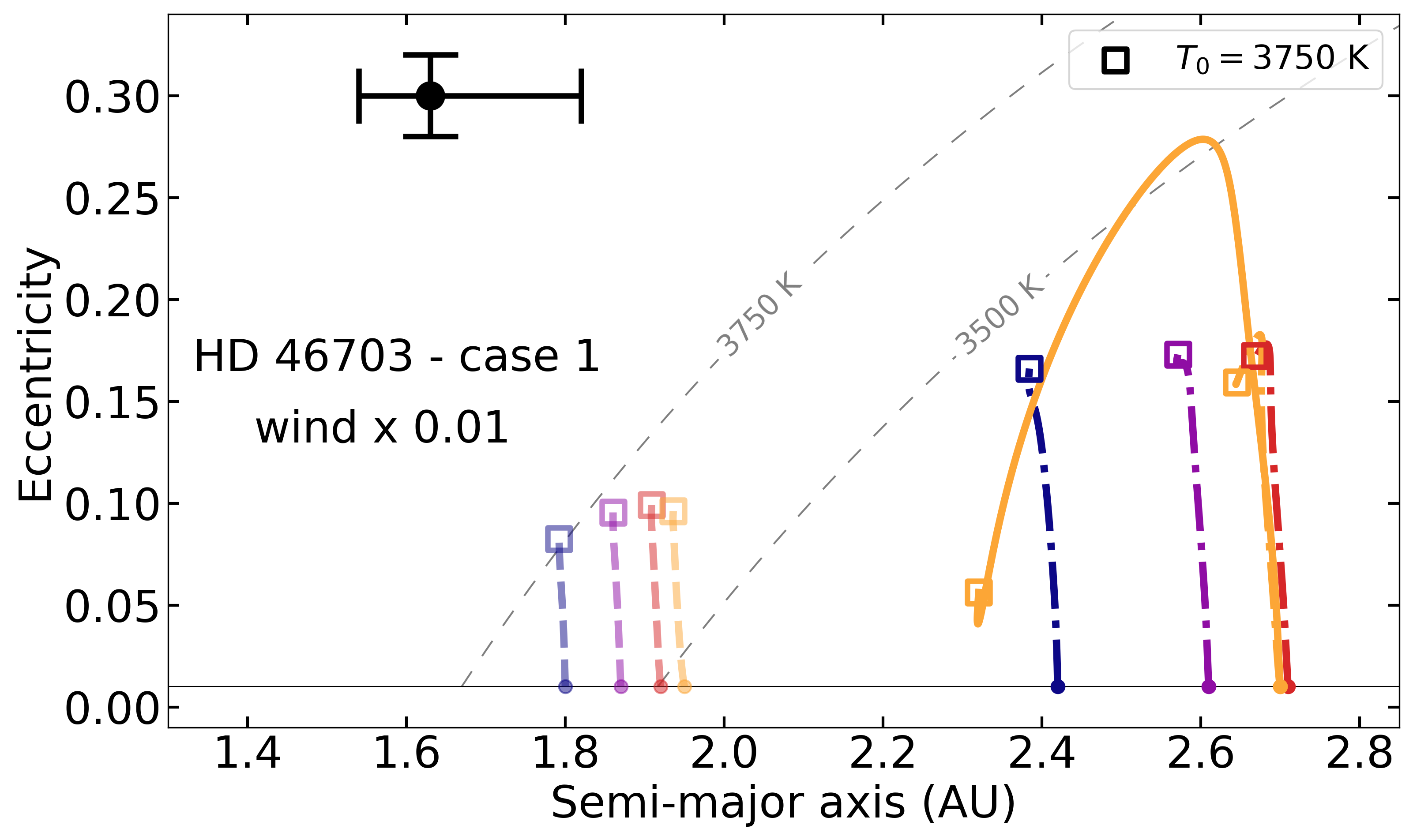}}
\caption{Eccentricity versus semi-major axis plot for case-1 models with $T_0 = 3750$~K. The four transparent models on the left have standard wind parameters (bottom panel of Fig.~\ref{fig:case1}), while the five models on the right are the successful models of Fig.~\ref{fig:hd46703_weakwind} with a 100 times weaker wind.}
\label{fig:hd46703case1ww}
\end{figure}

Stellar winds in post-AGB stars still pose a major uncertainty in their evolution \citep{cranmer11,millerbertolami16}. In paper~I we investigated the effect of stellar winds on the formation of depletion patterns; more precisely, we investigated its effect on the accretion properties required to reproduce the observed depletion patterns. We showed that weaker winds result in a much slower evolution of post-AGB stars. Therefore, higher disc masses are required for the star to become depleted.

In our standard model, we use a semi-empirical Reimers-type wind prescription as given by Eq.~4 of \citet{schroder05} for the wind mass-loss rate in the post-AGB phase. To evaluate the effect of different mass-loss rates on our results, we produced models with a weaker stellar wind for the case of HD~46703. We reduced the wind mass-loss rate by a factor of 100 and performed the depletion modelling and orbital evolution modelling as before. 

Figure~\ref{fig:hd46703_weakwind} shows the \texttt{MESA} models for HD~46703 in the $\log(\element{X_0}/\element{X})-T_\mathrm{eff}$ plane for models with $T_0 = 3750$~K in the weak-wind limit. Comparing with the standard model in the bottom panel of Fig.~\ref{fig:hd46703depletion}, the initial disc mass needs to be higher in order to explain the depletion pattern of HD~46703. This is due to the much longer evolutionary timescale of the star in the weak-wind regime, as can be seen in Table~\ref{table:accretionparams}. 

For the successful models, we plot the binary evolution for case~1 in Fig.~\ref{fig:hd46703case1ww}. Comparing these to the standard wind models, we find that the models require much larger initial semi-major axes. For the model with $M_\mathrm{d,0} = 10^{-1.5}$~$M_\sun$ (solid orange line), we see the large effect of tides at reducing the eccentricity. Because of accretion, the model gained envelope mass and expanded, such that $T_\mathrm{eff}$ dropped to 3500~K. Furthermore, because of the longer evolution timescale, tides are much stronger since the star remains large for a very long time. Consequently, all models are limited by the tidal interaction before the eccentricity reaches the observed value.

To summarise, the effect of a weaker wind on our results is a much slower evolution, which means higher initial disc masses are required to produce our depletion pattern. The latter results in stronger disc-binary interactions, while the former gives more time to pump the eccentricity. Despite this, we find that the models are still unable to explain the orbit for the case of HD~46703. As the eccentricity increases, the equilibrium tide becomes stronger and starts to counteract the pumping mechanisms. We conclude that, even though stellar winds play an important role in our results, the observed orbits of post-AGB binaries still cannot be explained.

\section{Discussion} \label{sect:discussion}


The goal of this work is to investigate whether disc-binary interactions can produce the eccentric orbits in post-AGB stars and other post-interaction binaries. Figures~\ref{fig:case1}--\ref{fig:case4} show that none of the evaluated interaction mechanisms provide a satisfactory solution for the orbits of EP~Lyr, RU~Cen, and HD~46703. In particular the accretion torque as given in \citet{munoz19a} does not significantly impact the eccentricity throughout post-AGB evolution. The effect of tides and resonances (if present) do play a relatively important role, but can only pump the eccentricity up to $\sim0.2$. 

\subsection{Orbital discrepancy between observations and models}

A key factor in the failure to reproduce the observed orbits is the evolution timescale of the stars. Models that start at higher temperatures evolve very rapidly to their current point in evolution. The eccentricity-pumping mechanisms therefore have insufficient time to produce the observed orbital eccentricities. Low-$T_0$ models generally reach higher eccentricities. These models evolve slowly, because they start their evolution with a high-mass envelope. They require high disc masses to become depleted, which increases the strength of the interactions. For these models, tides play an important role, since the stellar radius remains large for a long time. We find that as the eccentricity of those models increases, the torques evolve to an equilibrium situation in which $\dot{e}_\mathrm{tides} \approx - \dot{e}_\mathrm{res}$. Because the resonances start to cancel for increasing eccentricity while tides become much stronger, we expect this equilibrium to arise for all orbits as $e$ reaches $\sim 0.2$, unless the star spins up fast enough for tidal circularisation to become ineffective (Fig.~\ref{fig:corotation}), or unless the stellar radius has sufficiently decreased such that the tidal interaction becomes weaker.

Another important issue for explaining the orbit arises with the requirement for the star to remain within its Roche lobe throughout post-AGB evolution. Given their current luminosities and orbits, the stars in our sample will fill their Roche lobes at temperatures below about 4500~K. Post-AGB stars spend a long time at low $T_\mathrm{eff}$, but evolve quickly once the temperature increases beyond 4500~K. For that reason, the orbit does not change much when the star becomes hotter than 4500~K. In order to explain the orbit, one needs to pump the eccentricity during the slower stage in evolution, that is at low $T_\mathrm{eff}$ but large $R_*$. However, this conflicts with the requirement to avoid RLOF.

In nature, the stars could in principle evolve with an additional phase of RLOF. Even though we cannot include this in our current models, we expect this to change the models in two ways. Firstly, RLOF would manifest itself as an extra mass-loss agent for the post-AGB star, which speeds up the evolution of the star. Because the post-AGB phase is a phase of contraction and the companion is higher mass, mass transfer will always be stable. Secondly, the orbit will also change as a result of mass transfer. Phase-dependent RLOF could also have an eccentricity pumping effect \citep{soker00,bonacicmarinovic08}, but this would only have a small contribution because of the tiny amount of mass left in the envelope that can be transferred.

A further conundrum concerns the orbits of wide post-RGB systems like RU~Cen. Post-RGB stars must have formed via mass transfer, since stellar winds alone are not expected to be strong enough to remove the envelope on the RGB. The \M{present semi-major axis} of RU~Cen is too large to accommodate a Roche-lobe filling RGB star in a circular orbit. Therefore, we imposed a maximum initial semi-major axis of 3~AU. However, the disc-binary interactions we investigate primarily result in a decrease of $a_\mathrm{b}$, making it impossible to reproduce the orbit in this way. 

\begin{figure}
\resizebox{\hsize}{!}{\includegraphics{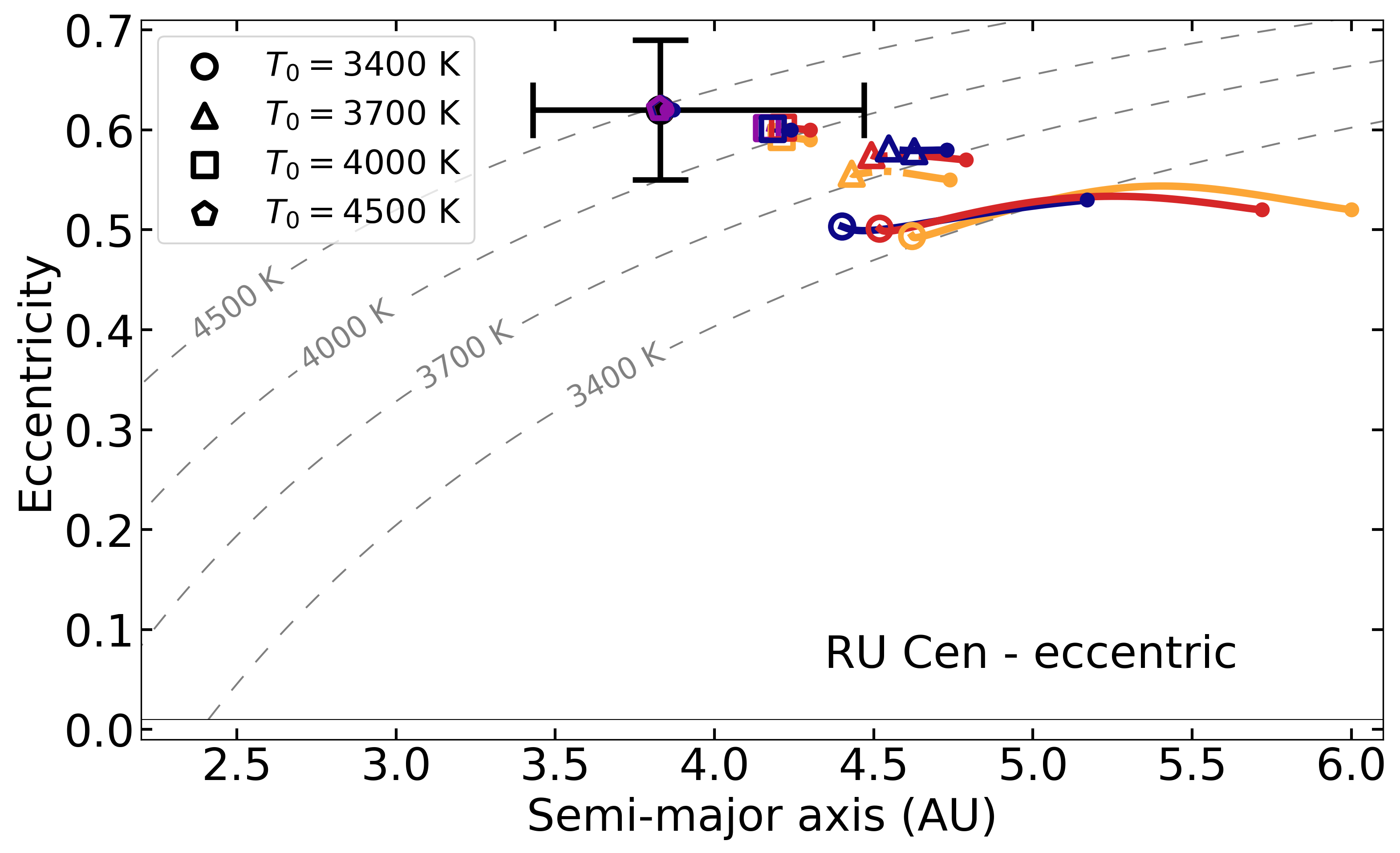}}
\caption{Eccentricity versus semi-major axis plot for case~1 models of RU~Cen. These models are similar to the middle panel of Fig.~\ref{fig:case1}, except the initial eccentricity is allowed to be non-zero. The star must remain inside its Roche lobe during its evolution. Starting points are given by a small circle, the end points are shown as open symbols. For $T_0=4500$~K the orbital parameters hardly change and coincide with the observed point.}
\label{fig:rucen_nonzero_e}
\end{figure}

This result suggests that this system already left the preceding phase of strong interaction in an eccentric orbit. \M{We show such models for the case of RU~Cen in Fig.~\ref{fig:rucen_nonzero_e}, where we allow for a non-zero initial eccentricity. The low-$T_0$ models are still somewhat limited by their Roche lobes. The current Roche-lobe size of RU~Cen at periastron passage suggests that the star detached with $T_\mathrm{eff} \approx 4500$~K. Models that have $T_0 = 4500$~K evolve on fast enough timescales such that they are not affected by tidal evolution or disc-binary interactions, hence their orbits remain essentially unchanged. Consequently, Fig.~\ref{fig:rucen_nonzero_e} shows that the orbit of RU~Cen, as well as its depletion pattern, can be reproduced if the initial orbit is allowed to be eccentric.} This formation scenario requires the eccentricity of the binary to have been set by the unknown binary interaction mechanism that led to the expulsion of the progenitor's envelope, while slightly widening the orbit. This can potentially be accomplished by mass loss and RLOF near periastron \citep{bonacicmarinovic08}, especially when the mass transfer rate becomes very large and the mass-loss timescale becomes much shorter than the circularisation timescale \citep[e.g.][]{kashi18}.

\M{Finally, other eccentricity pumping mechanisms during the post-AGB phase could be at play that we have not considered. A notable possibility could be dynamical interactions with a tertiary component \citep{toonen20}, in which the Lidov-Kozai effect can induce eccentricity in the inner and outer orbits \citep[see][and references therein]{naoz16}. Even though it is unlikely that the eccentric orbits of the whole population of post-AGB binaries are the result of triple interactions, it could provide a solution for a fraction of the systems, including special cases such as RU~Cen.}

\subsection{Model uncertainties}

It is important to note that the luminosity has a large impact on the results. The luminosity plays a dominant role in the evolution timescale of the star, since it determines both the nuclear burning rate and the stellar wind strength. Higher-luminosity stars have a much shorter evolution timescale, which is a key parameter in our models to pump the eccentricity. Moreover, luminosity determines the radius of a star at a given temperature. Lower-luminosity stars are therefore less constrained by their Roche-lobe sizes during their orbital evolution and experience a weaker tidal interaction. 

In this work, we used an empirical PLC relation to estimate the luminosity. The pulsation period is generally well-determined and is related to the radius of the star. For RV Tauri pulsators, there is a lot of intrinsic scatter in the PLC relation \citep{ripepi15}. The large pulsation periods also hamper the spectroscopic determination of $T_\mathrm{eff}$, which can vary over the orbital cycle by up to 1500~K. The changing colour of the star therefore induces additional uncertainty in deriving the luminosity from the PLC relation. Especially for HD~46703, the uncertainty is large due to the semi-regular nature of the pulsation period. It would therefore be preferable to use parallax-based luminosities.

Currently, the second data release of \textit{Gaia} does not take into account the effect of binary orbits on the parallax determination, hence this is too unreliable to use in this work. In particular orbits with periods in the range of 100--700 days are sensitive to systematic errors \citep{pourbaix19}. We look forward to the second part of \textit{Gaia} DR3, which will improve on the distances (and luminosities) significantly. This will allow for better constraints on the stellar model in addition to those of the PLC relation.

In addition to the luminosity, there are several uncertainties in the depletion analysis that propagate to our models. The values we derived for the dilution factor are sensitive to the \element{O} abundance, for which there are few good lines to use for the abundance determination. It is therefore difficult to trace back the abundances in the stellar atmosphere before the depletion process initiated. However, it can be seen in Figs.~\ref{fig:hd46703depletion}--\ref{fig:eplyrdepletion} that the accretion models are not very sensitive to the dilution factor, since the dilution of the envelope \M{in our models} increases steeply around 6000~K. \M{A small difference in observed dilution factor can therefore be modelled with similar accretion parameters.} An accurate determination of $T_\mathrm{eff}$ for these stars is however important.

For the modelling of depletion patterns in \texttt{MESA}, the stellar wind strength plays an important role. It has a large impact on the evolution timescale, hence on our results as was shown in Sect.~\ref{sect:assumptions}. Furthermore, the amount of mixing in the envelope of the star changes the rate at which the atmosphere becomes diluted, as discussed in paper~I. However, this effect is not expected to be important for temperatures below about 6000~K.

The disc model also introduces uncertainties in our results. In this work, we used a very simple model for the viscous evolution of the disc. As explained in Appendix~\ref{sect:discmodel}, our disc model assumes evolution without a central torque. However, we know that the disc-binary interactions investigated in this work exchange angular momentum between disc and binary. This impacts the temporal behaviour of the accretion rate, which might decrease more slowly with time than given by Eq.~\ref{eq:accroomen} (see Appendix~\ref{sect:discmodel}).

Furthermore, other disc evolution mechanisms could be at play other than viscosity. In particular disc mass loss can be effective in post-AGB binaries. A potentially important effect is photoevaporation, which is expected to be one of the leading mechanisms for the dispersal of protoplanetary discs \citep{alexander14,owen19}. High-energy photons, primarily produced by the accretion of gas onto the central stars, ionise the disc stripping it of gas and dust. This mechanism could be especially important for post-RGB models with longer evolutionary timescales, since accretion is expected to dominate early on in protoplanetary disc evolution \citep{alexander14}. Other possible mechanisms that could accelerate disc evolution are magnetohydrodynamical disc winds \citep{blandford82} and the formation of second-generation planets \citep{turner14}, although post-AGB discs are presumed to be stable against gravitational instabilities \citep{kluska18}.

\subsection{Additional model constraints}

For the complete sample of post-AGB binaries, it turns out that many orbits have an eccentricity in the range of 0.2--0.3 \citep[see Fig.~5 of][]{oomen18}. This is only slightly larger than the eccentricities we find in our models. It is therefore possible that some post-AGB orbits can be explained with disc-binary interactions, assuming that the resonance torque is indeed as strong as is claimed by \citet{lubow96}. It is still a concern that the strength of this torque was not reproduced in more recent works by for example \citet{munoz19a}.

In future work, we will expand our sample to other post-AGB binaries with saturated depletion patterns. For these objects, we only have lower limits on the required accretion rates and disc masses required to reproduce the depletion (see paper~I for an extensive discussion). Although this removes important constraints on the disc model and post-AGB evolution, such an analysis will provide additional insights in the general reproducibility of the orbits of post-AGB binaries by disc-binary interactions. Moreover, the sample of saturated-type post-AGB binaries is much larger and more diverse.

Observed disc properties could provide important additional constraints on the accretion models derived from the depletion analysis in \texttt{MESA}. For now, there is much degeneracy in the parameter space of $\dot{M}_0$, $M_\mathrm{d,0}$, and $T_0$. Deriving disc parameters, such as current disc mass, inner rim radius/temperature, and outer disc radius, could partly alleviate these degeneracies. However, this requires a detailed interferometric study of these systems. To probe the inner rim dust structure, near- or mid-infrared observations are required \citep[e.g.][]{kluska18, ertel19}, while the outer gas structure can be probed at mm-wavelengths \citep[e.g.][]{bujarrabal18}. \citet{kluska19} obtained near-IR data with PIONIER on the VLTI for RU~Cen, but could not detect the dust inner rim which is probably too cold. For EP~Lyr and HD~46703, currently no interferometric data is available.

\M{The disc masses of post-AGB binaries are very uncertain, even for well-studied systems such as IRAS~08544-4431. From a radiative transfer model of the SED and interferometric data, it is possible to obtain a rough estimate of the total dust mass in the disc \citep{kluska18}. However, the total disc mass is still subject to the uncertain gas-to-dust ratio of the disc. On the other hand, mm-wavelength studies can trace the total mass of the gas component at larger radii \citep{bujarrabal18}. This assumes that the CO transition lines are optically thin, which is not always the case.} Regardless, based on the small IR excesses in the SED of the objects in our sample, we estimate the current disc masses to be smaller than $\sim 10^{-3}~M_\sun$ for EP~Lyr and HD~46703. This favours models with initial disc masses of $10^{-3}~M_\sun$, which reach the current point in evolution with disc masses in the range of $3-8\times 10^{-4}~M_\sun$. However, these models reach relatively low eccentricities $< 0.05$. Another possibility is that the discs start out more massive, but lose mass due to extra disc mass-loss mechanisms that we did not take into account. 

In addition to the IR excess, a time series of optical spectra can give some information on the accretion properties. A jet launched by the companion as a result of its accretion produces a P-Cygni profile in the $\element{H}\alpha$ line when the companion moves in front of the post-AGB star, at its superior conjunction. Based on the jet absorption, it is possible to derive the accretion rate onto the companion \citep{bollen19}.

\section{Conclusions} \label{sect:conclusions}
We modelled the effect of accretion during the post-AGB phase with the \texttt{MESA} code to constrain the evolution of the star and circumbinary disc by reproducing the observed chemical depletion pattern. Using these constraints, we investigated several disc-binary interaction processes. We analysed the orbital evolution by including the effects of gas accretion inside the binary cavity, Lindblad resonances, and the equilibrium tide. We summarise our findings below.

We find that none of the models are able to simultaneously reproduce both depletion and orbit for the three stars in our sample (i.e. EP~Lyr, RU~Cen, and HD~46703). The different disc-binary interaction mechanisms we investigated cannot pump the eccentricity to the observed value within the evolutionary timescales of the stars. The torques that are produced by accretion streams and accretion discs inside the cavity of the circumbinary disc have very little effect on the binary eccentricity, and cause a small expansion of the orbit. The torque originating from the outer $(l,m)=(1,2)$ Lindblad resonance can efficiently pump the eccentricity to about 0.2, but quickly loses its potency at higher eccentricities. At that point, the equilibrium tide becomes of similar strength, inhibiting a further increase in $e$.

In addition to tidal interaction, the requirement to avoid RLOF is the main constraining factor. As the binary becomes more eccentric, the Roche-lobe radius becomes smaller around periastron, quickly leading to RLOF. In order to avoid RLOF during post-AGB evolution, the initial semi-major axis needs to be large enough such that the post-AGB star remains inside its Roche lobe at all times. Consequently, most models in which the eccentricity pumping starts at low $T_\mathrm{eff}$ and large $R_*$ end up in too wide orbits compared to the observed semi-major axis. 

Potentially successful models therefore need to start their post-AGB evolution at higher $T_\mathrm{eff}$. Since the stellar radius is smaller, the model can avoid RLOF and is less affected by tidal interaction. However, the evolution timescale at high $T_\mathrm{eff}$ becomes increasingly short, and the disc-binary interaction mechanisms do not have enough time to pump the eccentricity. Consequently, the general trend in our results is that models that start accretion at low starting temperatures ($T_0$) become more eccentric but have too wide orbits since they had to avoid RLOF, while models that start at high $T_0$ have a semi-major axis close to the observed value, but end with low eccentricity.

Within this framework, the orbits of wide post-RGB systems, such as RU~Cen, cannot be reproduced. Since post-RGB systems must be the result of mass transfer, \M{during} the interaction \M{on the RGB the orbit} must be small enough for RLOF and is expected to be circular \M{when the star enters the post-RGB phase}. Since the strongest disc-binary interactions (Lindblad resonances) have a tendency to decrease the semi-major axis, this can never lead to the wide, eccentric orbit of RU~Cen.

Another possible solution to explain the orbits of these systems is that they leave their previous binary interaction phase in an eccentric orbit. If the stars in our sample became detached while in their current orbit, the evolution occurs too fast for tides to change the orbit, while accretion can still produce the observed depletion pattern. However, this would require the system to have gained its eccentricity during the previous phase of strong mass loss which led to the expulsion of the envelope.

We investigated the impact of several assumptions on our results and found that these do not impact our results in such a way that the models provide a better fit to the observed orbits. There could be other factors limiting our results, such as the poorly known stellar luminosities or the simple disc model used in this work. We expect the former to improve with the third data release of \textit{Gaia}. The latter could still be improved by including a more detailed prescription of the effect of torques on disc evolution. Furthermore, effects such as photo-evaporation could become relevant for the long-lived post-RGB discs, although this would be an additional disc mass loss mechanism and result in weaker disc-binary interactions. Finally, we intend to expand our analysis to a larger sample of post-AGB stars in the future.

\begin{acknowledgements}
GMO acknowledges support of the Research Foundation - Flanders under contract G075916N and under grant number V434818N.
HvW acknowledges support from the Research Council of the KU Leuven under grant number C14/17/082.
This research has made use of the SIMBAD database,
operated at CDS, Strasbourg, France. This research has made use of NASA's Astrophysics Data System Bibliographic Services.
The authors would like to thank the anonymous referee for the constructive comments that have improved this paper. Furthermore, the authors thank Rob Izzard and Christoffel Waelkens for their additional feedback.
\end{acknowledgements}

\bibliographystyle{aa}
\bibliography{references.bib}

\begin{appendix}

\section{Disc model} \label{sect:discmodel}
The disc is described by a viscous evolution model. It can be shown that if the evolution of the disc is dominated by internal viscosity, a self-similar solution can be found for the evolution of the disc \citep{pringle91,rafikov16a}. This allows us to derive the time evolution of important disc properties, such as disc mass and accretion rate, based on their initial values.

In this work, we focus on the simple case in which viscosity is independent of the local surface density. A well-known solution occurs when there is no torque at the centre of the disc, such that the accretion rate can be described by \citep{rafikov16b}
\begin{equation}
\dot{M}(t) = \frac{M_\mathrm{d,0}}{2t_0}\left(1+\frac{t}{t_0}\right)^{-\frac{3}{2}},
\label{eq:accrrafikov}
\end{equation}
where $M_\mathrm{d,0}$ is the initial disc mass, and $t_0$ is the initial viscous timescale given by
\begin{equation}
t_0 = \frac{4}{3}\frac{\mu}{k_B}\frac{a_\mathrm{b}}{\alpha}\left(\frac{4\pi\sigma(GM_\mathrm{b})^2}{\zeta L_*}\right)^{1/4}\left(\frac{\eta}{I_L}\right)^2,
\label{eq:viscoustimescale}
\end{equation}
with $\mu$ the mean atomic weight, $k_B$ the Boltzmann constant, $a_\mathrm{b}$ the semi-major axis, $\alpha$ the viscosity parameter, $\sigma$ the constant of Stefan-Boltzmann, $M_\mathrm{b}$ the total mass of the binary, $L_*$ the luminosity of the post-AGB star, and $\zeta$ a constant that describes the geometrical effect of disc irradiation ($\sim0.1$). The two final parameters in Eq.~\ref{eq:viscoustimescale} are related to the angular momentum of the disc, with $\eta$ giving the specific angular momentum of the disc in units of specific binary angular momentum, and $I_L$ is a constant that describes the distribution of angular momentum \citep[$\approx 1.8$, see Fig.~5 of][for $\lambda = 0.5$]{rafikov16a}. 

The accretion rate given by Eq.~\ref{eq:accrrafikov} decreases over time as a result of the decreasing density in the disc. This is caused by two factors. Firstly, the total mass in the disc decreases as matter flows into the binary cavity. Secondly, as the disc evolves, the outer radius increases over time, thereby decreasing the overall density in the disc. Without a central torque, this evolution results in a time dependence of $\dot{M}\propto t^{-3/2}$. However, in this work, we aim to investigate the case of a non-zero central torque, which impacts the time-dependent behaviour of the disc \citep{rafikov16a, rafikov16b}. Recent 2D hydrodynamical models of finite discs suggest the accretion rate evolves as $\dot{M} \propto t^{-4/3}$ \citep{munoz19b}. However, in this work we will assume the simple form of accretion evolution with the formulation as given in Eq.~\ref{eq:accrrafikov}.

In the model of \citet{rafikov16a}, the presence of a non-zero central torque requires the inner surface density distribution to scale with $r^{-3/2}$. Even though observations of post-AGB stars find that a double power-law distribution better fits the density distribution \citep{hillen15, kluska18}, these observations are based on the dust distribution, which might be different from the distribution of gas. Consequently, we will keep the simple power-law distribution for the surface density with $\Sigma = Dr^{-3/2}$ for the sake of consistency with our model. 

The proportionality constant $D$ for the surface density profile is computed by requiring that integrating the surface density over the whole disc gives the total disc mass:
\begin{equation}
M_\mathrm{d} = \int^{R_\mathrm{out}}_{R_\mathrm{in}} \Sigma(r) 2\pi r \mathrm{d}r.
\label{eq:discmassint}
\end{equation}
To evaluate the integral in Eq.~\ref{eq:discmassint}, we need to determine the inner and outer boundaries of the disc. The size of the cavity becomes larger when the eccentricity increases \citep{mosta19} and depends on the structure of the disc \citep{artymowicz94}. The inner radius of the disc can be expressed as \citep{artymowicz94,dermine13}
\begin{equation}
R_\mathrm{in} = a_\mathrm{b}\left[1.7 + \frac{3}{8}\sqrt{e}\log\left(\alpha^{-1}\left(\frac{H}{R}\right)^{-2}\right)\right],
\label{eq:innerradius}
\end{equation}
where $(H/R)$ is the aspect ratio of the disc \M{which we assume to be equal to 0.2}. The outer radius of the disc $R_\mathrm{out}$ increases over time as a result of the viscous evolution of the disc. The initial value of the outer disc radius is a free parameter and will determine the initial angular momentum of the disc $J_\mathrm{d,0}$ since
\begin{equation}
J_\mathrm{d} = \int^{R_\mathrm{out}}_{R_\mathrm{in}} \Sigma(r) \Omega(r) 2\pi r^3 \mathrm{d}r.
\label{eq:angmomint}
\end{equation}
Here, we take $R_\mathrm{out,0} = 10 a_\mathrm{b}$, though this initial value is arbitrary. As the disc evolves, the outer radius can be derived by solving Eqs.~\ref{eq:discmassint} and \ref{eq:angmomint} for $R_\mathrm{out}(t)$. The integrals are evaluated numerically by dividing the radial coordinate into 1000 parts and applying the trapezium rule.

Similar to paper~I, we rewrite Eq.~\ref{eq:accrrafikov} as a function of the initial accretion rate $\dot{M}_0$ and initial disc mass $M_\mathrm{d,0}$ by substituting $t_0 = M_\mathrm{d,0}/(2\dot{M}_0)$, such that
\begin{equation}
\dot{M}(t) = \dot{M}_0\left(1+\frac{2\dot{M}_0t}{M_\mathrm{d,0}}\right)^{-3/2}.
\label{eq:accroomenapp}
\end{equation}
In Eq.~\ref{eq:accroomenapp}, $\dot{M}(t)$ is defined as the rate at which mass flows from the disc into the binary cavity. Since mass ratios in post-AGB binaries are in general larger than 0.3 \citep{oomen18}, both components of the binary system accrete roughly equal amounts of mass \citep{farris14, munoz19b, duffell19}. Consequently, we assume in our \texttt{MESA} models that the post-AGB star accretes half of the mass entering the cavity, or $\dot{M}_\mathrm{PAGB} = 0.5\times \dot{M}_\mathrm{bin}$.

\section{Equilibrium tide} \label{sect:equilibriumtide}
We include the effect of tides on the orbital evolution to properly investigate the evolution of the binary \citep{zahn77}. We follow \citet{hut81} for an (exact) formulation of the equilibrium tide. These tides are caused by the deformation of the star under the gravitational influence of the companion. This produces a tidal bulge that lags behind or precedes the motion of the secondary in the case of a rotational spin that is not synchronised to the orbit. More precisely, the bulge is offset with respect to the line connecting the centres of both stars, which produces a gravitational torque on the post-AGB star. 

This simple model of the equilibrium tide has been derived by \citet{hut81} and is given by
\begin{align}
\begin{split}
\dot{e} = -27\left(\frac{k}{T}\right)_\mathrm{c} q^{-1}(1+q^{-1})\left(\frac{R_*}{a_\mathrm{b}}\right)^8\frac{e}{(1-e^2)^{13/2}}\\
\times \left[f_3(e^2)-\frac{11}{18}(1-e^2)^{3/2} f_4(e^2)\frac{\Omega_\mathrm{spin}}{\Omega_\mathrm{orb}}\right],
\end{split}
\label{eq:tides}
\end{align}
where $f_3$ and $f_4$ are defined as in \citet{hut81}:
\begin{align}
f_3(e^2) &= 1 + \frac{15}{4}e^2 + \frac{15}{8}e^4 + \frac{5}{64}e^6\\
f_4(e^2) &= 1 + \frac{3}{2}e^2 + \frac{1}{8}e^4.
\end{align}
Here, $q$ is the mass ratio defined as $M_1/M_2$ and the ratio $\left(\frac{k}{T}\right)_\mathrm{c}$ is computed using Eq.~30 in \citet{hurley02}. 

The final term in Eq.~\ref{eq:tides} includes the ratio of the spin frequency over the orbital frequency, which determines the position of the tidal bulge with respect to the line of centres. If $\Omega_\mathrm{spin} > \frac{18}{11} \Omega_\mathrm{orb}$ (in the quasi-circular limit), the torque exerted by the bulge leads to a pumping of the eccentricity, while if $\Omega_\mathrm{spin} < \frac{18}{11} \Omega_\mathrm{orb}$, the eccentricity tends to decrease. 

For post-AGB stars, we expect the progenitor system to exit its phase of strong interaction in a synchronised fashion. Synchronicity does not necessarily persist during post-AGB evolution, hence we evaluate the change of rotation rate of the star as it evolves. Several processes act to change the rotation rate of the star. In what follows, we assume that the envelope is rigidly rotating and completely decoupled from the degenerate core.

First of all, the star shrinks and the envelope mass decreases as a result of stellar evolution. This causes the moment of inertia of the envelope to become smaller and the star to rotate faster. We compute the moment of inertia $I$ of the envelope at each time step in our \texttt{MESA} model by taking the sum $\sum \frac{2}{3}m_ir_i^2$ from the base of the envelope to the surface, where $m_i$ is the mass of cell $i$ and $r_i$ is the distance of that cell to the centre of the star. 

In addition to changes to the structure, the star can gain angular momentum due to accretion of gas from the circumbinary disc. We assume that the accreted gas has the specific angular momentum of Keplerian rotation at the surface of the star. Since gas will be accreted at the equator, we can write $\dot{J}_\mathrm{accr,s} = \dot{M}\sqrt{GM_*R_*}$, where $M_*$ and $R_*$ are the mass and radius of the post-AGB star. The value of $\dot{M}(t)$ is known at each time step from the disc model, while $R_*$ is given by our \texttt{MESA} model.

Even though accretion is taking place, stellar winds and nuclear burning cause the mass of the envelope to decrease over time. Stellar winds also take away angular momentum at the surface of the star. This is computed as $\dot{J}_\mathrm{wind,s} = \frac{2}{3}\dot{M}_\mathrm{wind}R_*^2\Omega_\mathrm{spin}$, where $\dot{M}_\mathrm{wind}$ is the mass lost from stellar winds and is a negative quantity. We assume that the wind is spherically symmetric, hence the factor of $2/3$.

Finally, tides drive the post-AGB star towards co-rotation with the orbit. Similar to the eccentricity evolution of Eq.~\ref{eq:tides}, we apply the formalism derived by \citet{hut81} such that $\dot{J}_\mathrm{tides,s}$ is given by
\begin{align}
\begin{split}
\dot{J}_\mathrm{tides,s} = 3\left(\frac{k}{T}\right)_\mathrm{c} \sqrt{G(M_1+M_2)a_\mathrm{b}} \frac{M_2^2}{M_1} \left(\frac{R_*}{a_\mathrm{b}}\right)^8 \frac{1}{(1-e^2)^{6}}\\
\times \left[f_2(e^2)-(1-e^2)^{3/2} f_5(e^2)\frac{\Omega_\mathrm{spin}}{\Omega_\mathrm{orb}}\right],
\end{split}
\label{eq:tidesspin}
\end{align}
where $f_2(e^2)$ and $f_5(e^2)$ are defined as
\begin{align}
f_2(e^2) &= 1 + \frac{15}{2}e^2 + \frac{45}{8}e^4 + \frac{5}{16}e^6\\
f_5(e^2) &= 1 + 3e^2 + \frac{3}{8}e^4.
\end{align}

\begin{figure}
\resizebox{\hsize}{!}{\includegraphics{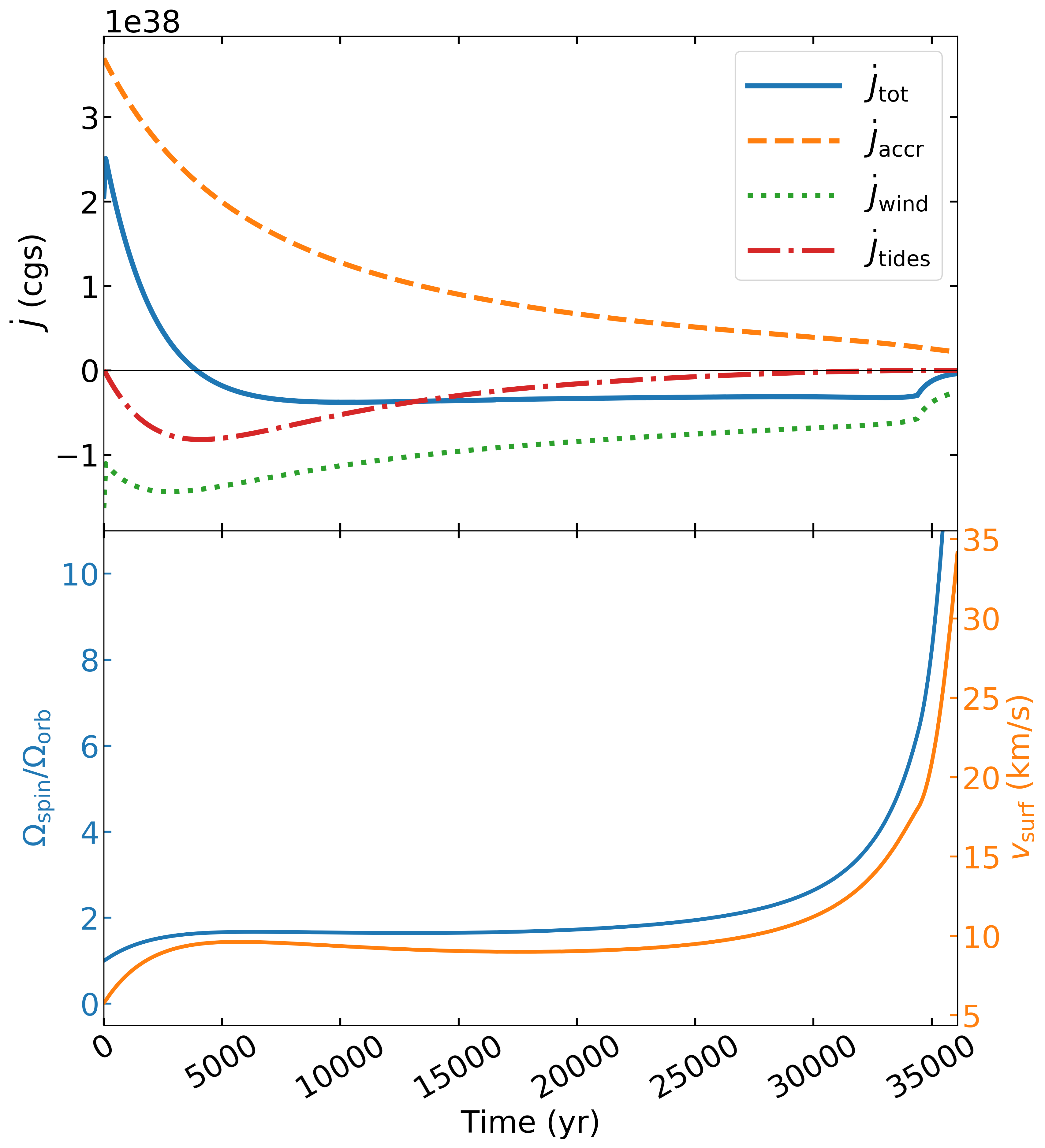}}
\caption{Spin evolution of HD~46703 for a model with $T_0 = 3500$~K, $\dot{M}_0 = 10^{-6.3}$~$M_\sun~\mathrm{yr}^{-1}$, and $M_\mathrm{d,0} = 10^{-2}$~$M_\sun$. \textit{Upper panel}: torques acting on the rotation of the post-AGB star due to accretion (dashed orange line), stellar winds (dotted green line), and tides (dash-dotted red line). \textit{Lower panel}: evolution of the surface rotational velocity at the equator (orange line). The blue line gives the ratio $\Omega_\mathrm{spin}/\Omega_\mathrm{orb}$, which starts at co-rotation ($=1$).}
\label{fig:corotation}
\end{figure}

Given these four processes, we compute the effect on the rotation of the post-AGB star with the initial condition of co-rotation with the orbit. The initial spin angular momentum of the envelope is given by $J_\mathrm{spin,0} = I_0\Omega_\mathrm{orb,0}$. The different torques acting on the envelope give
\begin{equation}
\dot{J}_\mathrm{spin} = \dot{J}_\mathrm{accr,s} + \dot{J}_\mathrm{wind,s} + \dot{J}_\mathrm{tides,s}.
\label{eq:torquesspin}
\end{equation}
The contraction of the star does not exert a torque, but changes the moment of inertia over time. At each time step, the new spin angular momentum and moment of inertia of the star determine the rotation rate as $\Omega_\mathrm{spin} = J_\mathrm{spin}/I$. Because the angular frequency of rotation changes, the effect of tides change as well (Eq.~\ref{eq:tides} and \ref{eq:tidesspin}). We show as an example the evolution of the rotation rate for a model of HD~46703 with $T_0 = 3500$~K in Fig.~\ref{fig:corotation}.

The upper panel of Fig.~\ref{fig:corotation} shows that early on in the evolution, accretion of gas increases the spin angular momentum, which quickly increases the angular velocity (lower panel). Because $\Omega_\mathrm{spin}/\Omega_\mathrm{orb}$ can become larger than $18/11$ in some cases, equilibrium tides can cause the eccentricity to increase rather than decrease. 

\M{In our implementation of tides in our simulations, we assume that tidal dissipation does not impact the envelope structure of the post-AGB star. Indeed, the energy dissipation rate is a small fraction of a solar luminosity, hence is several orders of magnitude smaller than the stellar luminosity.}

\section{Model results} \label{sect:modelresults}
\M{In Table~\ref{table:accretionparams}, we present the parameters of the successful accretion models for each star from Figs.~\ref{fig:hd46703depletion}--\ref{fig:eplyrdepletion}. We label each individual model in the second column. Models H1--H14 are the different models of HD~46703, while H15WW--H19WW are models where we assume a weaker wind. For RU~Cen, the models are labelled R1--R15 and for EP~Lyr these are labelled E1--E14. In the next three columns, we show the parameter combinations of the successful models, which are given by the starting temperature $T_0$, the initial accretion rate $\dot{M}_0$ in units of $M_\sun~\mathrm{yr}^{-1}$, and the initial disc mass $M_\mathrm{d,0}$ in units of $M_\sun$. Columns 5--7 give the evolution timescale in years and the disc properties at the end of the evolution (i.e. the predicted current values), with $\dot{M}_\mathrm{f}$ being the final accretion rate onto the binary and $M_\mathrm{d,f}$ the final disc mass of the model.}

\M{In Tables~\ref{table:resultshd46703}--\ref{table:resultseplyr}, we present the results of the binary evolution calculations. We have omitted models for which RLOF during the evolution could not be avoided. The first column shows the method of the calculations as presented in Table~\ref{table:cases}, while in the second column we give the labelled accretion models from Table~\ref{table:accretionparams}. In the remaining columns, we have from left to right the initial semi-major axis $a_\mathrm{b,0}$ in AU, the final semi-major axis $a_\mathrm{b,f}$ in AU, the final eccentricity $e_\mathrm{f}$, the final outer disc radius $R_\mathrm{out}$ in AU, and the viscosity parameter $\alpha$ which was used in the model. The viscosity parameter was calculated from Eq.~\ref{eq:viscoustimescale} by using the accretion properties and the initial semi-major axis.}

\begin{table*}
\caption{Initial and final parameters of the accretion model for each of the stars in the sample.}
\centering
\begin{tabular}{c c c c c c c c}
\hline\hline
Star & \# & $T_0$ (K) & $\dot{M}_0$ & $M_\mathrm{d,0}$& $t_\mathrm{evol}$& $\dot{M}_\mathrm{f}$& $M_\mathrm{d,f}$ \\
\hline
\multirow{19}{*}{HD~46703}
& H1 & 3250 & $10^{-7.0}$ & $10^{-1.5}$ & 116\,000 & $4.38\times10^{-8}$ & $2.4\times10^{-2}$ \\
& H2 & 3250 & $10^{-6.3}$ & $10^{-1.5}$ & 134\,000 & $4.17\times10^{-8}$ & $1.38\times10^{-2}$ \\
& H3 & 3250 & $10^{-6.0}$ & $10^{-1.5}$ & 141\,000 & $3.22\times10^{-8}$ & $1.01\times10^{-2}$ \\
& H4 & 3500 & $10^{-7.0}$ & $10^{-2.0}$ & 30\,000 & $4.90\times10^{-8}$ & $7.89\times10^{-3}$ \\
& H5 & 3500 & $10^{-6.3}$ & $10^{-2.0}$ & 36\,000 & $5.05\times10^{-8}$ & $4.65\times10^{-3}$ \\
& H6 & 3500 & $10^{-6.0}$ & $10^{-2.0}$ & 38\,000 & $3.95\times10^{-8}$ & $3.41\times10^{-3}$ \\
& H7 & 3750 & $10^{-7.0}$ & $10^{-2.5}$ & 14\,000 & $3.94\times10^{-8}$ & $2.32\times10^{-3}$ \\
& H8 & 3750 & $10^{-6.6}$ & $10^{-2.5}$ & 15\,000 & $4.09\times10^{-8}$ & $1.73\times10^{-3}$ \\
& H9 & 3750 & $10^{-6.3}$ & $10^{-2.5}$ & 16\,000 & $3.43\times10^{-8}$ & $1.29\times10^{-3}$ \\
& H10 & 3750 & $10^{-6.0}$ & $10^{-2.5}$ & 17\,000 & $2.58\times10^{-8}$ & $9.35\times10^{-4}$ \\
& H11 & 4500 & $10^{-7.0}$ & $10^{-3.0}$ & 3500 & $4.52\times10^{-8}$ & $7.67\times10^{-4}$ \\
& H12 & 4500 & $10^{-6.6}$ & $10^{-3.0}$ & 3900 & $4.98\times10^{-8}$ & $5.83\times10^{-4}$ \\
& H13 & 4500 & $10^{-6.3}$ & $10^{-3.0}$ & 4200 & $4.23\times10^{-8}$ & $4.39\times10^{-4}$ \\
& H14 & 4500 & $10^{-6.0}$ & $10^{-3.0}$ & 4700 & $3.02\times10^{-8}$ & $3.11\times10^{-4}$ \\
& H15WW & 3750 & $10^{-7.0}$ & $10^{-2.0}$ & 163\,000 & $1.13\times10^{-8}$ & $4.84\times10^{-3}$ \\
& H16WW & 3750 & $10^{-6.6}$ & $10^{-2.0}$ & 191\,000 & $7.27\times10^{-9}$ & $3.07\times10^{-3}$ \\
& H17WW & 3750 & $10^{-6.3}$ & $10^{-2.0}$ & 206\,000 & $4.99\times10^{-9}$ & $2.15\times10^{-3}$ \\
& H18WW & 3750 & $10^{-6.0}$ & $10^{-2.0}$ & 216\,000 & $3.41\times10^{-9}$ & $1.51\times10^{-3}$ \\
& H19WW & 3750 & $10^{-6.0}$ & $10^{-1.5}$ & 531\,000 & $4.91\times10^{-9}$ & $5.38\times10^{-3}$ \\
\hline
\multirow{15}{*}{RU~Cen}
& R1 & 3400 & $10^{-7.0}$ & $10^{-1.5}$ & 80\,000 & $5.42\times10^{-8}$ & $2.58\times10^{-2}$ \\
& R2 & 3400 & $10^{-6.3}$ & $10^{-1.5}$ & 105\,000 & $5.55\times10^{-8}$ & $1.52\times10^{-2}$ \\
& R3 & 3400 & $10^{-6.0}$ & $10^{-1.5}$ & 114\,000 & $4.23\times10^{-8}$ & $1.1\times10^{-2}$ \\
& R4 & 3700 & $10^{-7.0}$ & $10^{-2.5}$ & 24\,000 & $2.53\times10^{-8}$ & $2.0\times10^{-3}$ \\
& R5 & 3700 & $10^{-7.0}$ & $10^{-2.0}$ & 25\,000 & $5.39\times10^{-8}$ & $8.14\times10^{-3}$ \\
& R6 & 3700 & $10^{-6.3}$ & $10^{-2.0}$ & 36\,000 & $5.14\times10^{-8}$ & $4.68\times10^{-3}$ \\
& R7 & 3700 & $10^{-6.0}$ & $10^{-2.0}$ & 39\,000 & $3.86\times10^{-8}$ & $3.38\times10^{-3}$ \\
& R8 & 4000 & $10^{-7.0}$ & $10^{-2.5}$ & 11\,000 & $4.48\times10^{-8}$ & $2.42\times10^{-3}$ \\
& R9 & 4000 & $10^{-6.6}$ & $10^{-2.5}$ & 13\,000 & $4.63\times10^{-8}$ & $1.8\times10^{-3}$ \\
& R10 & 4000 & $10^{-6.3}$ & $10^{-2.5}$ & 15\,000 & $3.75\times10^{-8}$ & $1.33\times10^{-3}$ \\
& R11 & 4000 & $10^{-6.0}$ & $10^{-2.5}$ & 16\,000 & $2.67\times10^{-8}$ & $9.45\times10^{-4}$ \\
& R12 & 4500 & $10^{-7.0}$ & $10^{-3.0}$ & 4800 & $3.63\times10^{-8}$ & $7.13\times10^{-4}$ \\
& R13 & 4500 & $10^{-7.0}$ & $10^{-2.5}$ & 5100 & $6.58\times10^{-8}$ & $2.75\times10^{-3}$ \\
& R14 & 4500 & $10^{-6.6}$ & $10^{-3.0}$ & 5500 & $3.46\times10^{-8}$ & $5.16\times10^{-4}$ \\
& R15 & 4500 & $10^{-6.3}$ & $10^{-3.0}$ & 6200 & $2.59\times10^{-8}$ & $3.73\times10^{-4}$ \\
\hline
\multirow{15}{*}{EP~Lyr}
& E1 & 3500 & $10^{-6.6}$ & $10^{-1.5}$ & 7700 & $2.11\times10^{-7}$ & $2.98\times10^{-2}$ \\
& E2 & 3500 & $10^{-6.3}$ & $10^{-2.0}$ & 8000 & $2.07\times10^{-7}$ & $7.45\times10^{-3}$ \\
& E3 & 3500 & $10^{-6.3}$ & $10^{-1.5}$ & 8400 & $3.52\times10^{-7}$ & $2.81\times10^{-2}$ \\
& E4 & 3500 & $10^{-6.0}$ & $10^{-2.0}$ & 8500 & $2.26\times10^{-7}$ & $6.09\times10^{-3}$ \\
& E5 & 4000 & $10^{-6.6}$ & $10^{-2.0}$ & 2700 & $2.08\times10^{-7}$ & $9.39\times10^{-3}$ \\
& E6 & 4000 & $10^{-6.6}$ & $10^{-1.5}$ & 2700 & $2.36\times10^{-7}$ & $3.1\times10^{-2}$ \\
& E7 & 4000 & $10^{-6.3}$ & $10^{-2.5}$ & 2800 & $1.95\times10^{-7}$ & $2.31\times10^{-3}$ \\
& E8 & 4000 & $10^{-6.3}$ & $10^{-2.0}$ & 2900 & $3.42\times10^{-7}$ & $8.8\times10^{-3}$ \\
& E9 & 4000 & $10^{-6.0}$ & $10^{-2.5}$ & 2900 & $2.07\times10^{-7}$ & $1.87\times10^{-3}$ \\
& E10 & 4500 & $10^{-6.6}$ & $10^{-2.5}$ & 1300 & $1.90\times10^{-7}$ & $2.88\times10^{-3}$ \\
& E11 & 4500 & $10^{-6.6}$ & $10^{-2.0}$ & 1300 & $2.28\times10^{-7}$ & $9.68\times10^{-3}$ \\
& E12 & 4500 & $10^{-6.6}$ & $10^{-1.5}$ & 1300 & $2.44\times10^{-7}$ & $3.13\times10^{-2}$ \\
& E13 & 4500 & $10^{-6.3}$ & $10^{-2.5}$ & 1400 & $2.89\times10^{-7}$ & $2.63\times10^{-3}$ \\
& E14 & 4500 & $10^{-6.3}$ & $10^{-2.0}$ & 1500 & $4.06\times10^{-7}$ & $9.32\times10^{-3}$ \\
& E15 & 4500 & $10^{-6.0}$ & $10^{-2.5}$ & 1600 & $3.43\times10^{-7}$ & $2.21\times10^{-3}$ \\
\hline
\end{tabular}
\label{table:accretionparams}
\end{table*}

\longtab[2]{
\begin{longtable}{c c c c c c c}
\caption{Results of orbital evolution models for cases 1, 2, 3, 4, and for testing assumptions for HD~46703.\label{table:resultshd46703}}\\
\hline\hline
Method & \# & $a_\mathrm{b,0}$ & $a_\mathrm{b,f}$ & $e_\mathrm{f}$ & $R_\mathrm{out}$ & $\alpha$ \\
\hline
\endfirsthead
\caption{continued.}\\
\hline\hline
Method & \# & $a_\mathrm{b,0}$ & $a_\mathrm{b,f}$ & $e_\mathrm{f}$ & $R_\mathrm{out}$ & $\alpha$ \\
\hline
\endhead

\endfoot

\multirow{14}{*}{Case~1}
& H1 & 2.7 & 2.52 & 0.14 & 90 & $9.9\times10^{-4}$ \\
& H2 & 2.7 & 2.37 & 0.10 & 410 & $5.\times10^{-3}$ \\
& H3 & 2.7 & 2.34 & 0.08 & 850 & $9.9\times10^{-3}$ \\
& H4 & 2.24 & 2.21 & 0.12 & 70 & $2.6\times10^{-3}$ \\
& H5 & 2.44 & 2.4 & 0.16 & 320 & $1.4\times10^{-2}$ \\
& H6 & 2.53 & 2.49 & 0.17 & 680 & $2.9\times10^{-2}$ \\
& H7 & 1.8 & 1.79 & 0.08 & 70 & $6.6\times10^{-3}$ \\
& H8 & 1.87 & 1.86 & 0.10 & 170 & $1.7\times10^{-2}$ \\
& H9 & 1.92 & 1.91 & 0.10 & 360 & $3.5\times10^{-2}$ \\
& H10 & 1.95 & 1.94 & 0.10 & 780 & $7.2\times10^{-2}$ \\
& H11 & 1.59 & 1.59 & 0.04 & 60 & $1.8\times10^{-2}$ \\
& H12 & 1.56 & 1.56 & 0.05 & 130 & $4.6\times10^{-2}$ \\
& H13 & 1.54 & 1.53 & 0.04 & 260 & $9.\times10^{-2}$ \\
& H14 & 1.54 & 1.53 & 0.03 & 580 & $1.8\times10^{-1}$ \\
\hline

\multirow{14}{*}{Case~2}
& H1 & 2.7 & 2.44 & 0.13 & 100 & $9.9\times10^{-4}$ \\
& H2 & 2.7 & 2.23 & 0.08 & 500 & $5.\times10^{-3}$ \\
& H3 & 2.7 & 2.18 & 0.06 & 1060 & $9.9\times10^{-3}$ \\
& H4 & 2.25 & 2.2 & 0.12 & 70 & $2.6\times10^{-3}$ \\
& H5 & 2.49 & 2.4 & 0.15 & 390 & $1.5\times10^{-2}$ \\
& H6 & 2.58 & 2.47 & 0.15 & 860 & $3.\times10^{-2}$ \\
& H7 & 1.81 & 1.8 & 0.08 & 80 & $6.7\times10^{-3}$ \\
& H8 & 1.88 & 1.86 & 0.09 & 210 & $1.7\times10^{-2}$ \\
& H9 & 1.92 & 1.9 & 0.09 & 440 & $3.5\times10^{-2}$ \\
& H10 & 1.95 & 1.92 & 0.09 & 970 & $7.2\times10^{-2}$ \\
& H11 & 1.59 & 1.59 & 0.04 & 60 & $1.8\times10^{-2}$ \\
& H12 & 1.56 & 1.55 & 0.04 & 150 & $4.6\times10^{-2}$ \\
& H13 & 1.54 & 1.53 & 0.04 & 320 & $9.\times10^{-2}$ \\
& H14 & 1.56 & 1.55 & 0.03 & 720 & $1.8\times10^{-1}$ \\
\hline
\multirow{14}{*}{Case~3}
& H1 & 2.13 & 2.2 & 0.01 & 40 & $7.8\times10^{-4}$ \\
& H2 & 2.13 & 2.31 & 0.01 & 110 & $3.9\times10^{-3}$ \\
& H3 & 2.13 & 2.36 & 0.01 & 200 & $7.8\times10^{-3}$ \\
& H4 & 1.91 & 1.93 & 0.01 & 30 & $2.2\times10^{-3}$ \\
& H5 & 1.91 & 1.96 & 0.01 & 90 & $1.1\times10^{-2}$ \\
& H6 & 1.91 & 1.97 & 0.01 & 160 & $2.2\times10^{-2}$ \\
& H7 & 1.66 & 1.67 & 0.01 & 30 & $6.1\times10^{-3}$ \\
& H8 & 1.66 & 1.67 & 0.01 & 50 & $1.5\times10^{-2}$ \\
& H9 & 1.81 & 1.83 & 0.01 & 110 & $3.3\times10^{-2}$ \\
& H10 & 1.82 & 1.84 & 0.01 & 200 & $6.7\times10^{-2}$ \\
& H11 & 1.63 & 1.63 & 0.01 & 30 & $1.9\times10^{-2}$ \\
& H12 & 1.63 & 1.63 & 0.01 & 50 & $4.8\times10^{-2}$ \\
& H13 & 1.63 & 1.63 & 0.01 & 80 & $9.5\times10^{-2}$ \\
& H14 & 1.63 & 1.64 & 0.01 & 160 & $1.9\times10^{-1}$ \\
\hline
\multirow{14}{*}{Case~4}
& H1 & 2.54 & 2.47 & 0.26 & 80 & $9.3\times10^{-4}$ \\
& H4 & 2.0 & 1.98 & 0.14 & 60 & $2.3\times10^{-3}$ \\
& H5 & 2.19 & 2.17 & 0.19 & 290 & $1.3\times10^{-2}$ \\
& H6 & 2.27 & 2.25 & 0.20 & 630 & $2.6\times10^{-2}$ \\
& H7 & 1.68 & 1.67 & 0.09 & 70 & $6.2\times10^{-3}$ \\
& H8 & 1.72 & 1.71 & 0.10 & 160 & $1.6\times10^{-2}$ \\
& H9 & 1.76 & 1.75 & 0.11 & 340 & $3.2\times10^{-2}$ \\
& H10 & 1.79 & 1.78 & 0.11 & 730 & $6.6\times10^{-2}$ \\
& H11 & 1.59 & 1.59 & 0.04 & 60 & $1.8\times10^{-2}$ \\
& H12 & 1.56 & 1.56 & 0.05 & 130 & $4.6\times10^{-2}$ \\
& H13 & 1.54 & 1.53 & 0.04 & 260 & $9.\times10^{-2}$ \\
& H14 & 1.54 & 1.53 & 0.03 & 580 & $1.8\times10^{-1}$ \\
\hline
\pagebreak
\multirow{4}{*}{$R_\mathrm{out,0} = 5a_\mathrm{b}$}
& H7 & 1.82 & 1.81 & 0.08 & 50 & $4.2\times10^{-3}$ \\
& H8 & 1.89 & 1.88 & 0.10 & 120 & $1.1\times10^{-2}$ \\
& H9 & 1.94 & 1.93 & 0.10 & 260 & $2.3\times10^{-2}$ \\
& H10 & 1.98 & 1.97 & 0.10 & 550 & $4.6\times10^{-2}$ \\
\hline
\multirow{4}{*}{$R_\mathrm{out,0} = 50a_\mathrm{b}$}
& H7 & 1.74 & 1.73 & 0.07 & 230 & $2.2\times10^{-2}$ \\
& H8 & 1.79 & 1.78 & 0.08 & 500 & $5.8\times10^{-2}$ \\
& H9 & 1.81 & 1.79 & 0.08 & 980 & $1.2\times10^{-1}$ \\
& H10 & 1.81 & 1.79 & 0.06 & 2020 & $2.3\times10^{-1}$ \\
\hline
\multirow{4}{*}{$\left(\frac{H}{R}\right) = 0.1$}
& H7 & 1.74 & 1.74 & 0.05 & 40 & $6.4\times10^{-3}$ \\
& H8 & 1.79 & 1.79 & 0.06 & 90 & $1.7\times10^{-2}$ \\
& H9 & 1.83 & 1.83 & 0.06 & 180 & $3.4\times10^{-2}$ \\
& H10 & 1.84 & 1.84 & 0.05 & 370 & $6.8\times10^{-2}$ \\
\hline
\multirow{5}{*}{Weak wind}
& H15WW & 2.42 & 2.38 & 0.17 & 290 & $2.8\times10^{-3}$ \\
& H16WW & 2.61 & 2.57 & 0.17 & 880 & $7.6\times10^{-3}$ \\
& H17WW & 2.71 & 2.67 & 0.17 & 1980 & $1.6\times10^{-2}$ \\
& H18WW & 2.7 & 2.65 & 0.16 & 4210 & $3.1\times10^{-2}$ \\
& H19WW & 2.7 & 2.32 & 0.06 & 3330 & $9.9\times10^{-3}$ \\
\hline

\end{longtable}
}

\longtab[3]{

\begin{longtable}{c c c c c c c}
\caption{Results of orbital evolution models for cases 1 and 4 for RU~Cen.\label{table:resultsrucen}}\\
\hline\hline
Method & \# & $a_\mathrm{b,0}$ & $a_\mathrm{b,f}$ & $e_\mathrm{f}$ & $R_\mathrm{out}$ & $\alpha$ \\
\hline
\endfirsthead
\caption{continued.}\\
\hline\hline
Method & \# & $a_\mathrm{b,0}$ & $a_\mathrm{b,f}$ & $e_\mathrm{f}$ & $R_\mathrm{out}$ & $\alpha$ \\
\hline
\endhead

\endfoot

\hline
\multirow{15}{*}{Case~1}
& R1 & 3.0 & 2.76 & 0.21 & 100 & $1.7\times10^{-3}$ \\
& R4 & 3.0 & 2.98 & 0.10 & 250 & $1.7\times10^{-2}$ \\
& R5 & 3.0 & 2.95 & 0.15 & 100 & $5.3\times10^{-3}$ \\
& R6 & 3.0 & 2.93 & 0.21 & 560 & $2.7\times10^{-2}$ \\
& R7 & 3.0 & 2.92 & 0.21 & 1230 & $5.3\times10^{-2}$ \\
& R8 & 3.0 & 2.98 & 0.09 & 130 & $1.7\times10^{-2}$ \\
& R9 & 3.0 & 2.97 & 0.10 & 340 & $4.2\times10^{-2}$ \\
& R10 & 3.0 & 2.97 & 0.10 & 730 & $8.4\times10^{-2}$ \\
& R11 & 3.0 & 2.96 & 0.09 & 1610 & $1.7\times10^{-1}$ \\
& R12 & 3.0 & 2.99 & 0.04 & 170 & $5.3\times10^{-2}$ \\
& R13 & 3.0 & 2.99 & 0.07 & 80 & $1.7\times10^{-2}$ \\
& R14 & 3.0 & 2.98 & 0.03 & 440 & $1.3\times10^{-1}$ \\
& R15 & 3.0 & 2.98 & 0.02 & 970 & $2.7\times10^{-1}$ \\
\hline
\multirow{15}{*}{Case~4}
& R1 & 3.0 & 2.85 & 0.27 & 100 & $1.7\times10^{-3}$ \\
& R4 & 3.0 & 2.98 & 0.10 & 250 & $1.7\times10^{-2}$ \\
& R5 & 3.0 & 2.95 & 0.15 & 100 & $5.3\times10^{-3}$ \\
& R6 & 3.0 & 2.93 & 0.21 & 560 & $2.7\times10^{-2}$ \\
& R7 & 3.0 & 2.92 & 0.22 & 1220 & $5.3\times10^{-2}$ \\
& R8 & 3.0 & 2.98 & 0.09 & 130 & $1.7\times10^{-2}$ \\
& R9 & 3.0 & 2.97 & 0.10 & 340 & $4.2\times10^{-2}$ \\
& R10 & 3.0 & 2.97 & 0.10 & 730 & $8.4\times10^{-2}$ \\
& R11 & 3.0 & 2.96 & 0.09 & 1610 & $1.7\times10^{-1}$ \\
& R12 & 3.0 & 2.99 & 0.04 & 170 & $5.3\times10^{-2}$ \\
& R13 & 3.0 & 2.99 & 0.07 & 80 & $1.7\times10^{-2}$ \\
& R14 & 3.0 & 2.98 & 0.03 & 440 & $1.3\times10^{-1}$ \\
& R15 & 3.0 & 2.98 & 0.02 & 970 & $2.7\times10^{-1}$ \\
\hline
\end{longtable}
}

\longtab[4]{

\begin{longtable}{c c c c c c c}
\caption{Results of orbital evolution models for cases 1 and 4 for EP~Lyr.\label{table:resultseplyr}}\\
\hline\hline
Method & \# & $a_\mathrm{b,0}$ & $a_\mathrm{b,f}$ & $e_\mathrm{f}$ & $R_\mathrm{out}$ & $\alpha$ \\
\hline
\endfirsthead
\caption{continued.}\\
\hline\hline
Method & \# & $a_\mathrm{b,0}$ & $a_\mathrm{b,f}$ & $e_\mathrm{f}$ & $R_\mathrm{out}$ & $\alpha$ \\
\hline
\endhead

\endfoot

\multirow{15}{*}{Case~1}
& E1 & 2.94 & 2.92 & 0.11 & 40 & $2.4\times10^{-3}$ \\
& E2 & 2.98 & 2.96 & 0.12 & 100 & $1.5\times10^{-2}$ \\
& E3 & 3.0 & 2.96 & 0.15 & 50 & $4.9\times10^{-3}$ \\
& E4 & 3.05 & 3.02 & 0.13 & 180 & $3.1\times10^{-2}$ \\
& E5 & 2.66 & 2.65 & 0.07 & 40 & $6.8\times10^{-3}$ \\
& E6 & 2.66 & 2.65 & 0.08 & 30 & $2.2\times10^{-3}$ \\
& E7 & 2.65 & 2.64 & 0.06 & 100 & $4.3\times10^{-2}$ \\
& E8 & 2.66 & 2.65 & 0.09 & 50 & $1.4\times10^{-2}$ \\
& E9 & 2.64 & 2.63 & 0.06 & 180 & $8.5\times10^{-2}$ \\
& E10 & 2.66 & 2.66 & 0.04 & 40 & $2.2\times10^{-2}$ \\
& E11 & 2.66 & 2.66 & 0.05 & 30 & $6.8\times10^{-3}$ \\
& E12 & 2.66 & 2.66 & 0.05 & 30 & $2.2\times10^{-3}$ \\
& E13 & 2.65 & 2.64 & 0.05 & 60 & $4.3\times10^{-2}$ \\
& E14 & 2.66 & 2.65 & 0.07 & 40 & $1.4\times10^{-2}$ \\
& E15 & 2.64 & 2.63 & 0.06 & 110 & $8.5\times10^{-2}$ \\
\hline
\multirow{15}{*}{Case~4}
& E1 & 2.88 & 2.86 & 0.12 & 40 & $2.3\times10^{-3}$ \\
& E2 & 2.93 & 2.91 & 0.12 & 100 & $1.5\times10^{-2}$ \\
& E3 & 2.95 & 2.92 & 0.16 & 50 & $4.8\times10^{-3}$ \\
& E4 & 3.0 & 2.98 & 0.14 & 180 & $3.1\times10^{-2}$ \\
& E5 & 2.66 & 2.65 & 0.07 & 40 & $6.8\times10^{-3}$ \\
& E6 & 2.66 & 2.65 & 0.08 & 30 & $2.2\times10^{-3}$ \\
& E7 & 2.65 & 2.64 & 0.06 & 100 & $4.3\times10^{-2}$ \\
& E8 & 2.66 & 2.65 & 0.09 & 50 & $1.4\times10^{-2}$ \\
& E9 & 2.64 & 2.63 & 0.06 & 180 & $8.5\times10^{-2}$ \\
& E10 & 2.66 & 2.66 & 0.04 & 40 & $2.2\times10^{-2}$ \\
& E11 & 2.66 & 2.66 & 0.05 & 30 & $6.8\times10^{-3}$ \\
& E12 & 2.66 & 2.66 & 0.05 & 30 & $2.2\times10^{-3}$ \\
& E13 & 2.65 & 2.64 & 0.05 & 60 & $4.3\times10^{-2}$ \\
& E14 & 2.66 & 2.65 & 0.07 & 40 & $1.4\times10^{-2}$ \\
& E15 & 2.64 & 2.63 & 0.06 & 110 & $8.5\times10^{-2}$ \\
\hline
\end{longtable}
}

\end{appendix}

\end{document}